\documentclass[twocolumn]{aastex631}
\usepackage{comment}
\usepackage{empheq}
\usepackage{hyperref}

\newcommand{\Add}[1]{#1}

\shorttitle{Understanding Prominence Eruption through Data-Driven Modeling and Magnetic Environment}
\shortauthors{Namekata et al.}

\graphicspath{{./}{../analysis-2022/figures/}{figures/}}

\begin{document}

\title{Multiwavelength Campaign Observations of a Young Solar-type Star, EK Draconis. II. Understanding Prominence Eruption through Data-Driven Modeling and Observed Magnetic Environment}


\author[0000-0002-1297-9485]{Kosuke Namekata}
\affiliation{The Hakubi Center for Advanced Research, Kyoto University, Kyoto 606-8302, Japan}
\affiliation{Department of Physics, Kyoto University, Kitashirakawa-Oiwake-cho, Sakyo-ku, Kyoto, 606-8502, Japan}
\affil{NASA Goddard Space Flight Center, 8800 Greenbelt Road, Greenbelt, MD 20771, USA}
\affiliation{Department of Physics, American University, Washington, DC, USA}
\affiliation{Division of Science, National Astronomical Observatory of Japan, NINS, Osawa, Mitaka, Tokyo, 181-8588, Japan}
\email{namekata@kusastro.kyoto-u.ac.jp}

\author[0000-0002-5978-057X]{Kai Ikuta}
\affil{Department of Multidisciplinary Sciences, The University of Tokyo, 3-8-1 Komaba, Meguro, Tokyo 153-8902, Japan}

\author[0000-0001-7624-9222]{Pascal Petit}
\affiliation{Institut de Recherche en Astrophysique et Plan\'{e}tologie, Universit\'{e} de Toulouse, CNRS, CNES, 14 avenue \'{E}douard Belin, 31400 Toulouse, France}

\author[0000-0003-4452-0588]{Vladimir S. Airapetian}
\affiliation{Sellers Exoplanetary Environments Collaboration, NASA Goddard Space Flight Center, Greenbelt, MD, USA}
\affiliation{Department of Physics, American University, Washington, DC, USA}

\author[0000-0001-5371-2675]{Aline A. Vidotto}
\affil{Leiden Observatory, Leiden University, PO Box 9513, 2300 RA Leiden, The Netherlands}

\author[0000-0002-5778-2600]{Petr Heinzel}
\affiliation{Center of Excellence `Solar and Stellar Activity', University of Wroc{\l}aw, Kopernika 11, PL-51622 Wroc{\l}aw, Poland}
\affiliation{Astronomical Institute of Czech Academy of Sciences Fri\v{c}ova 298, 251 65, Ond\v{r}ejov, Czech Republic}

\author[0000-0002-7618-9394]{Ji\v{r}\'{i} Wollmann}
\affiliation{Astronomical Institute of Czech Academy of Sciences Fri\v{c}ova 298, 251 65, Ond\v{r}ejov, Czech Republic}

\author[0000-0003-0332-0811]{Hiroyuki Maehara}
\affil{Okayama Branch Office, Subaru Telescope, National Astronomical Observatory of Japan, NINS, Kamogata, Asakuchi, Okayama 719-0232, Japan}

\author[0000-0002-0412-0849]{Yuta Notsu}
\affil{Laboratory for Atmospheric and Space Physics, University of Colorado Boulder, 3665 Discovery Drive, Boulder, CO 80303, USA}
\affil{National Solar Observatory, 3665 Discovery Drive, Boulder, CO 80303, USA}

\author[0000-0003-3085-304X]{Shun Inoue}
\affil{Department of Physics, Kyoto University, Sakyo, Kyoto 606-8502, Japan}

\author[0000-0001-5522-8887]{Stephen Marsden}
\affil{Centre for Astrophysics, University of Southern Queensland, Toowoomba, Queensland 4350, Australia}

\author[0000-0002-4996-6901]{Julien Morin}
\affil{LUPM, Universit\'e de Montpellier, CNRS, Place Eug\`ene Bataillon, F-34095 Montpellier, France}

\author[0000-0003-2490-4779]{Sandra V. Jeffers}
\affil{Max Planck Institute for Solar System Research, Justus-von-Liebig-weg 3, 37077 G\"ottingen, Germany}

\author[0000-0003-1978-9809]{Coralie Neiner}
\affil{LESIA, Paris Observatory, PSL University, CNRS, Sorbonne University, Universit\'e Paris Cit\'e, 5 place Jules Janssen, 92195 Meudon, France}


\author[0000-0002-8090-3570]{Rishi R. Paudel}
\affil{NASA Goddard Space Flight Center, 8800 Greenbelt Road, Greenbelt, MD 20771, USA}
\affil{University of Maryland, Baltimore County, 1000 Hilltop Circle, Baltimore, MD 21250, USA}

\author[0009-0001-5099-8070]{Antoaneta A. Avramova-Boncheva}
\affil{The Institute of Astronomy and National Astronomical Observatory,  Bulgarian Academy of Sciences, 72 Tsarigradsko Chaussee Blvd.,1784 Sofia, Bulgaria}

\author[0000-0001-7115-2819]{Keith Gendreau}
\affil{NASA Goddard Space Flight Center, Greenbelt, MD, USA}

\author{Kazunari Shibata}
\affil{Kwasan Observatory, Kyoto University, Yamashina, Kyoto 607-8471, Japan}
\affil{School of Science and Engineering, Doshisha University, Kyotanabe, Kyoto 610-0321, Japan}

\begin{abstract}

EK Draconis, a nearby young solar-type star (G1.5V, 50-120 Myr), is known as one of the best proxies for inferring the environmental conditions of the young Sun. 
The star frequently produces superflares and \textsf{Paper \hyperref[2022ApJ...926L...5N]{I}} presented the first evidence of an associated gigantic prominence eruption observed as a blueshifted H$\alpha$ Balmer line emission. 
In this paper, we present the results of dynamical modeling of the stellar eruption and examine its relationship to the surface starspots and large-scale magnetic fields observed concurrently with the event.
By performing a one-dimensional free-fall dynamical model and a one dimensional hydrodynamic simulation of the flow along the expanding magnetic loop, we found that the prominence eruption likely occurred near the stellar limb (12$^{+5}_{-5}$-16$^{+7}_{-7}$ degrees from the limb) and was ejected at an angle of 15$^{+6}_{-5}$-24$^{+6}_{-6}$ degrees relative to the line of sight, and the magnetic structures can expand into a coronal mass ejection (CME). 
\Add{The observed prominence displayed a terminal velocity of $\sim$0 km s$^{-1}$ prior to disappearance, complicating the interpretation of its dynamics in \textsf{Paper \hyperref[2022ApJ...926L...5N]{I}}.
The models in this paper suggest that prominence's H$\alpha$ intensity diminishes at around or before its expected maximum height, explaining the puzzling time evolution in observations.}
The TESS light curve modeling and (Zeeman) Doppler Imaging revealed large mid-latitude spots with polarity inversion lines and one polar spot with dominant single polarity, all near the stellar limb during the eruption. 
This suggests that mid-latitude spots could be the source of the pre-existing gigantic prominence we reported in \textsf{Paper \hyperref[2022ApJ...926L...5N]{I}}.
These results provide valuable insights into the dynamic processes that likely influenced the environments of early Earth, Mars, Venus, and young exoplanets.

\end{abstract}



\section{Introduction}\label{sec:1}

Solar and stellar flares are among the most powerful phenomena in the atmospheres of cool stars, observed across a wide range of wavelengths from radio to X-ray bands \citep{2011LRSP....8....6S,2017LRSP...14....2B}. These flares result from the conversion of magnetic energy into kinetic and thermal energy via magnetic reconnection. 
The most powerful solar flares have energies around 10$^{32}$ erg \citep{2012ApJ...759...71E}, while cosmogenic radionuclide studies \citep{2019esps.book.....M,2023LRSP...20....2U} and observations of Sun-like stars \citep{2012Natur.485..478M,2013ApJS..209....5S,2019ApJ...876...58N,2020arXiv201102117O} suggest potential ``superflares" ($>10^{33}$ erg) on our Sun.
On the Sun, large flares often accompany coronal mass ejections (CMEs), which eject massive amounts of coronal material into space at high speeds, impacting planetary magnetospheres and ionospheres \citep[e.g.,][]{2021LRSP...18....4T}.
Consequently, massive flares are garnering attention in terms of significant potential impacts on human society \citep[e.g.,][]{2022LRSP...19....2C}.

Motivated by the recent successful \Add{detections} of large \Add{numbers} of exoplanets, there has been an increased focus on the space weather environments of other stellar systems \citep[e.g.,][]{2019LNP...955.....L}. 
Numerous attempts have been made to estimate the complete X-ray and ultraviolet radiation \citep[e.g.,][]{2014ApJ...780...61L,2019A&A...624L..10J,2023ApJ...945..147N} and plasma environments \citep[e.g.,][]{2005ApJ...628L.143W} around stars; however, characterizing these environments is often challenging because stars cannot be spatially resolved and the signals of astrospheres are too weak.
Particularly, stellar CMEs associated with transient stellar flares present significant difficulties in detection and characterization of their propagation \citep[see reviews by][]{2017IAUS..328..243O,2022SerAJ.205....1L,2022arXiv221105506N}. 
Historically, observations of solar flares have guided both direct and indirect attempts to detect stellar CMEs. 
For example, numerous blueshifted emission profiles in chromospheric lines have been observed during flares from M/K-dwarfs and active close binaries, which are interpreted as a signature of stellar prominence eruptions, the lower parts or even cores of CMEs \citep{1990A&A...238..249H,1992AJ....104.1161E,1994A&A...285..489G,2006A&A...452..987C,2008A&A...487..293F,2011A&A...534A.133F,2018A&A...615A..14F,2016A&A...590A..11V,2019A&A...623A..49V,2017A&A...597A.101F,2018PASJ...70...62H,2020arXiv201200786K,2020A&A...637A..13M,2020MNRAS.499.5047M,2020PASJ..tmp..253M,2023ApJ...948....9I,2024ApJ...961..189N,2024PASJ...76..175I}. 
Though velocities are often low, some are fast enough to suggest prominence eruptions leading to CMEs \citep{1990A&A...238..249H,2016A&A...590A..11V,2019A&A...623A..49V,2023ApJ...948....9I}. 
\Add{Additionally, post-flare coronal dimming \citep{2021NatAs...5..697V,2022ApJ...936..170L} and type-IV radio bursts \citep{2020ApJ...905...23Z,2024A&A...686A..51M} have been reported to infer the occurrence of stellar CMEs mainly from M/K dwarfs.} 
Current reports lack crucial details such as type-II radio bursts \citep{2018ApJ...856...39C,2018ApJ...862..113C,2019ApJ...871..214V}, which indicate CME driven shock formation and propagation.
Therefore, the consensus in \Add{the} stellar community is that direct evidence of stellar CME occurrence has not yet been reported.
Numerical simulations assuming large-scale dipole magnetic fields suggest that a strong overlying coronal magnetic field may suppress eruptions \citep{2018ApJ...862...93A}, indicating that stellar CMEs can be rare or their signal is too weak compared to the solar case.
Comparing observations and theoretical models is important to reveal \Add{whether and how stellar eruptions happen}, yet there is a notable lack of simultaneous datasets of stellar CME signatures and magnetic field mappings, and consequently, the connection between theoretical predictions and observational data remains suboptimal.

Previous attempts to detect stellar CMEs \Add{have} been biased towards relatively cooler stars (e.g., M-dwarfs), with minimal research conducted on solar-type stars (G-type main-sequence stars). 
However, observing solar-type stars is crucial as it provides a significant window into the past and future evolution of the Sun and solar system planets \citep{1997ApJ...483..947G,2007LRSP....4....3G}. 
Previous observations of young solar-type stars suggest that the young Sun was highly active, possessed large surface magnetic flux \citep{2014MNRAS.441.2361V,2016MNRAS.457..580F,2018MNRAS.474.4956F,2020A&A...635A.142K}, and frequently emitted superflares \citep{1999ApJ...513L..53A,2000ApJ...541..396A,2022ApJ...926L...5N,2022PASJ...74.1295Y,2022A&A...661A.148C}.
It is proposed that, if these superflares are frequently associated with massive CMEs, the high-energy particles from these CMEs could have significantly altered the chemical composition of planets, potentially generating greenhouse gases, prebiotic chemistry, and complex biological molecules. \citep{2016NatGe...9..452A,2023Life...13.1103K}. 
In addition, the eruptive events could play a significant role in stellar mass and angular momentum loss evolution, which influences \Add{estimates of} how active the Sun and stars were during their youth \citep[e.g.,][]{2015ApJ...809...79O,2017ApJ...840..114C,2021ApJ...915...37W}.
Therefore, investigating the occurrence of CMEs from young solar-type stars is an essential task that relates to fundamental questions about our origins.

With the aim of investigating signals of superflares and filament/prominence eruptions as indirect evidence of CMEs,
we have conducted long-term spectroscopic monitoring observations of a young solar-type star, EK Draconis (EK Dra, HD 129333).
EK Dra is a G1.5V dwarf with the age of 50--125 Myr, the effective temperature of 5560--5700 K, radius of 0.94 $R_{\odot}$, and mass of 0.95 $M_{\odot}$ \citep{2017MNRAS.465.2076W,2021MNRAS.502.3343S}.
It is a rapidly rotating star with the rotation period of $\sim$2.77 days \citep{1999ApJ...513L..53A,2015AJ....150....7A,2022NatAs...6..241N}, and it exhibits a high level of magnetic activity in the form of \Add{a} hot and dense corona \citep[e.g.,][]{2005A&A...432..671S,2012ApJ...745...25L,2023ApJ...945..147N}, strong surface magnetic field and large starspots \citep[e.g.,][]{2005AN....326..283B,2017MNRAS.465.2076W,2018A&A...620A.162J} and frequent superflares \citep[e.g.,][]{1999ApJ...513L..53A,2000ApJ...541..396A,2015AJ....150....7A,2022NatAs...6..241N,2022ApJ...926L...5N}.
These \Add{characteristics} make this star one of the best targets to detect superflares and associated CMEs from an infant Sun at the time of the Hadean period on Earth.

As a result of our long-term monitoring observations since 2020, we have successfully detected the H$\alpha$ line spectra of five superflares on EK Dra \citep{2022NatAs...6..241N,2022ApJ...926L...5N,2024ApJ...961...23N}. 
Among these, one event exhibited a blue-shifted absorption component \citep[][later, followed up by \citealt{2024MNRAS.tmp.1385L}]{2022NatAs...6..241N}, and two were accompanied by blue-shifted emission components (\citealt{2024ApJ...961...23N}, hereafter referred to as \textsf{Paper \hyperref[2022ApJ...926L...5N]{I}}). 
Comparative analysis with solar observations suggests that the absorption profiles indicate filament eruptions occurring in front of the stellar disk, while the emission profiles imply prominence eruptions outside the stellar disk. 
Also, their high line-of-sight (LOS) velocities of 330--690 km s$^{-1}$, close to the escape velocity of $\sim$670 km s$^{-1}$, suggest the possible occurrence of stellar CMEs.
This suggestion is supported by a \Add{recent} one dimensional (1D) hydrodynamic (HD) simulation of \Add{the plasma flow along an expanding magnetic loop for the filament eruption} \citep{2024ApJ...963...50I}.
\Add{This discovery provides indirect evidence of stellar CMEs, marking the first indications specifically for young solar-type stars.}

Among these discoveries, the massive prominence eruption on April 10, 2022 (labeled ``E1" in \textsf{Paper \hyperref[2022ApJ...926L...5N]{I}}) presents particularly intriguing and well-characterized observational properties.
This event shows \Add{a} very massive eruptive prominence with a mass of $\sim$10$^{20}$ g and possible X-ray coronal dimming occurrence, providing the first multi-wavelength characterization of stellar CME signatures.
One of the intriguing aspects \Add{of this event} is that the velocity of the eruption began near the escape speed at 690 km s$^{-1}$, decelerated approximately along with surface gravity, and showed a terminal speed of $\sim$0 km s$^{-1}$ before it disappeared without a dominant redshift (unlike the filament eruption event ``E4," which eventually showed a redshift; \citealt{2022NatAs...6..241N}). 
\Add{This puzzling absence of redshift after zero velocity leads} to questions about how this eruptive prominence evolved.
Indeed, there are various strong observational constraints concerning this event:  
First, the fact that it was always visible in emission indicates that it was consistently outside the stellar disk. 
Secondly, the deceleration rate close to the surface gravity of the star suggests that it was launched from near the stellar surface, aligned somewhat along the line of sight (LOS). 
These factors constrain the possible dynamics significantly, leading us to use a simple numerical model to extract more detailed dynamics of this event, as will be shown in Section \ref{sec:3}.
Additionally, as mentioned in \textsf{Paper \hyperref[2022ApJ...926L...5N]{I}}, simultaneous spectropolarimetric observations at the T\'{e}lescope Bernard Lyot (TBL) and photometric observations by \Add{the} Transiting Exoplanet Survey Satellite (TESS, \citealt{2015JATIS...1a4003R}) were conducted. 
The spectropolarimetric observations at TBL have been used for Doppler Imaging (DI) and Zeeman Doppler Imaging (ZDI), enabling us to reconstruct the spot and magnetic field configurations.
Moreover, the TESS light curve shows quasi-periodic variations due to rotation with large starspots, which allows us to extract information \Add{on} starspot sizes and locations. 
In summary, we have obtained an unprecedented dataset that will help us infer the direction, evolution and origin of stellar prominence \Add{eruptions}.

Leveraging this dataset, in this study, we examine the dynamics of the prominence and its relationship to starspots and the global magnetic field geometry.
In Section \ref{sec:2}, we perform the spot and magnetic field mappings by using the above datasets from TBL (Section \ref{section:2-1}) and TESS (Section \ref{section:2-2}). 
In particular, we get independent intensity maps from both data which are compared with each other (Section \ref{sec:4-1}).
In Section \ref{sec:3}, we perform simple modelings of the dynamics of the prominence eruptions, based on the above strong observational constraints (Section \ref{sec:3-1}). 
We apply the simplest one dimensional free-fall model to the observation (Section \ref{sec:3-2}), and later apply a 1D HD simulation of the flow along the expanding magnetic loop (\Add{pseudo two-dimensional magnetohydrodynamics (2D MHD) model}, see Section \ref{sec:3-3}).
Finally, we discuss the dynamics and the relationship between prominence eruptions and surface mapping in Section \ref{sec:4}.

\section{Observations and Models of Starspot and Magnetic Field}\label{sec:2}

\subsection{Doppler and Zeeman Doppler Imaging}\label{section:2-1}

Spectropolarimetric observations of EK~Dra were recorded at Pic du Midi Observatory, using the Neo-NARVAL instrument \citep{2022A&A...661A..91L}. The time series consists of 18 observations collected in 2022, between March 22 and April 25.  From this data set, we discarded two observations suffering from a \Add{low} signal-to-noise ratio (S/N hereafter), using 16 observations in our analysis. All data were automatically reduced using the default data reduction software of the instrument. The reduced material obtained for each polarimetric sequence results in one intensity (Stokes I) and one circularly polarized (Stokes V) spectrum, covering the whole wavelength domain between 380 and 1,050~nm. We noticed that the bluest spectral orders of our observations were plagued by a very low S/N (a similar situation was reported by \cite{2023AJ....166..167M}, using the same instrumental setup). We, therefore, conservatively decided to ignore in every reduced observation all spectral bins bluer than 470~nm. 

\begin{deluxetable}{cccccc}
\label{tab:obs}
\tablecaption{The summary of spectropolarimetric observations of EK Dra using the TBL.}
\tablewidth{0pt}
\tablehead{
\colhead{UT} & \colhead{UT} & \colhead{JD} \\ 
\colhead{date} & \colhead{middle} & \colhead{middle}  
}
\startdata
2022 Mar 22  &  21:16:41  &  2459661.3866 \\ 
2022 Mar 23  &  04:32:22  &  2459661.6892 \\ 
2022 Mar 23  &  21:15:32  &  2459662.3858 \\ 
2022 Mar 24  &  04:44:13  &  2459662.6974 \\ 
2022 Mar 26  &  21:55:15  &  2459665.4134 \\ 
2022 Mar 28  &  04:03:55  &  2459666.6694 \\ 
2022 Apr 04  &  21:36:01  &  2459674.4000 \\ 
2022 Apr 05  &  04:05:26  &  2459674.6705 \\ 
2022 Apr 15  &  21:40:01  &  2459685.4028 \\ 
2022 Apr 16  &  21:58:57  &  2459686.4159 \\ 
2022 Apr 17  &  03:47:05  &  2459686.6577 \\ 
2022 Apr 17  &  22:01:35  &  2459687.4178 \\ 
2022 Apr 18  &  22:02:39  &  2459688.4185 \\ 
2022 Apr 19  &  03:50:18  &  2459688.6599 \\ 
2022 Apr 25  &  22:31:06  &  2459695.4383 \\ 
2022 Apr 26  &  03:53:10  &  2459695.6619 \\ 
\enddata
\end{deluxetable}

\begin{figure}
\centering
\epsscale{1.0}
\plotone{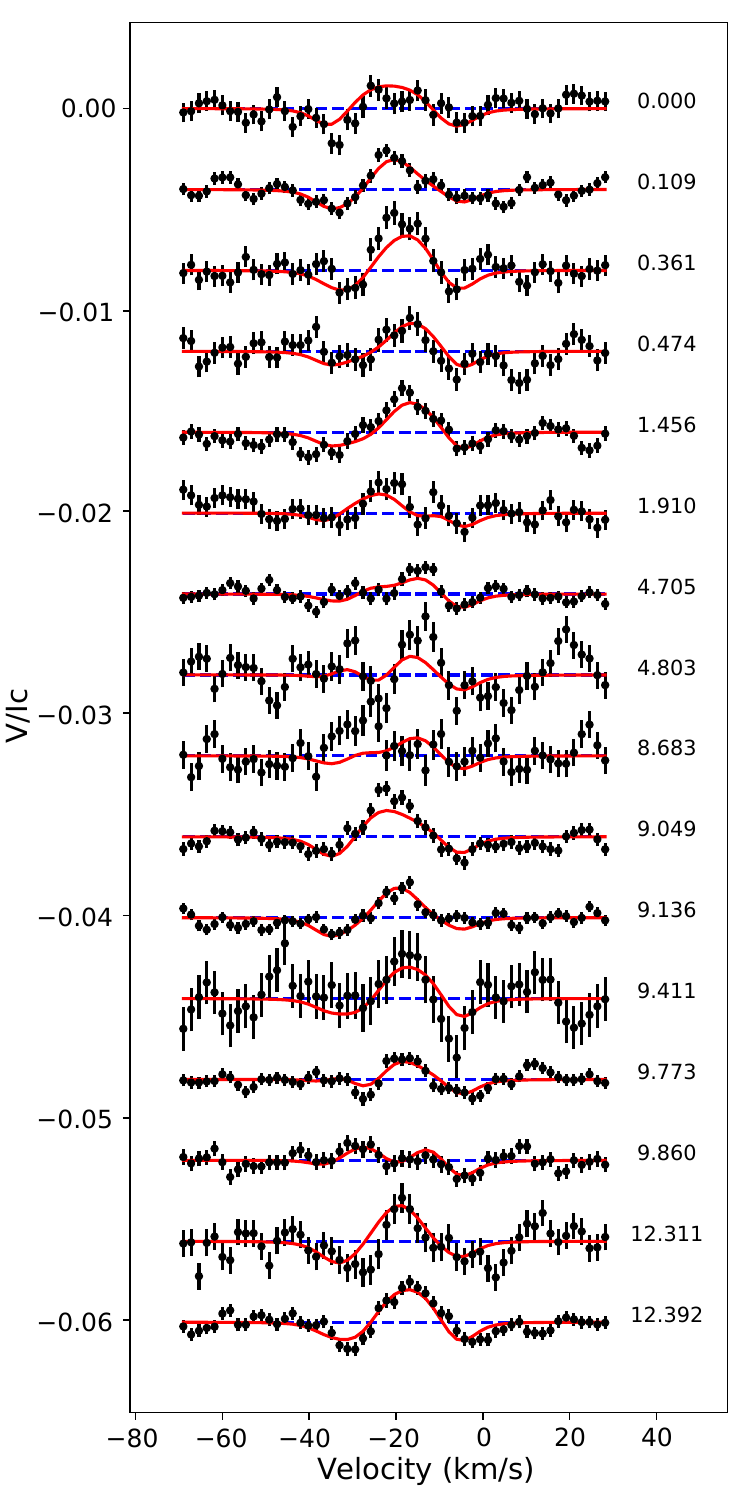}
 \caption{Observed Stokes V LSD profiles (back dots), and synthetic LSD profiles produced by the ZDI model (red curves). The blue dashed lines show the zero level. The numbers on the right indicate the rotation cycles. Successive observations are vertically shifted for plot clarity.}
\label{fig:lsdV}%
\end{figure}

\begin{figure*}
\centering
\epsscale{0.75}
\plotone{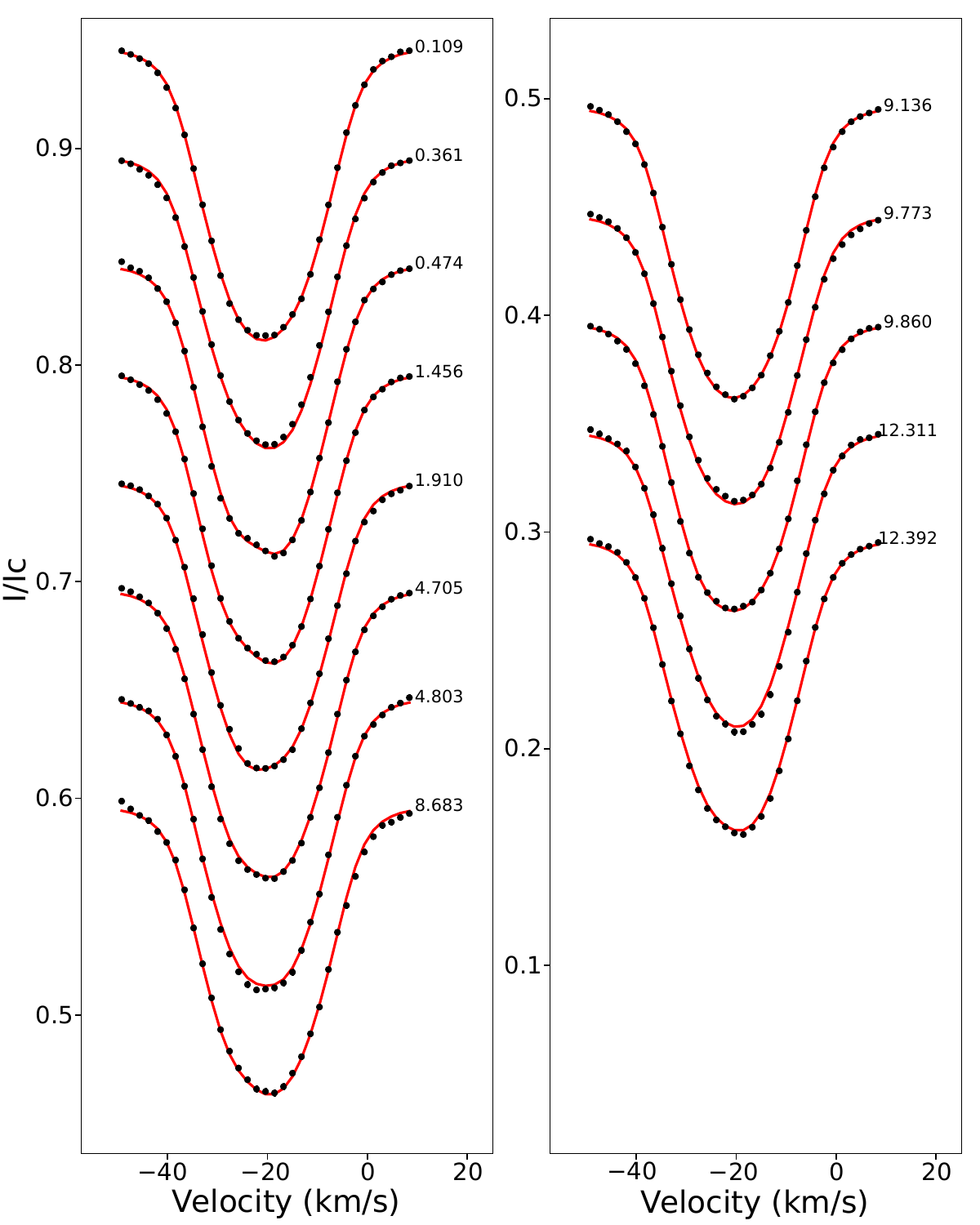}
\caption{Same as Figure \ref{fig:lsdV}, but for the Stokes I LSD profiles.}
\label{fig:lsdI}%
\end{figure*}

As generally done to detect weak, circularly polarized Zeeman signatures dominated by the noise, we used the Least-Squares Deconvolution method (LSD, \citealt{1997MNRAS.291..658D, 2010A&A...524A...5K}) to extract, from each spectrum, an average line profile in Stokes I and V. To do so, we used a list of photospheric spectral lines produced by the VALD database \citep{2015PhyS...90e4005R}, using a surface effective temperature and gravity close to the fundamental parameters of EK~Dra. We kept for our analysis atomic lines deeper than 40\% of the continuum level. We discarded lines in telluric bands and photospheric lines blended with strong chromospheric lines, resulting in 1,400 spectral lines for the LSD analysis. The sets of LSD line profiles are displayed in Figure \ref{fig:lsdV} and \ref{fig:lsdI}. For most observations, circularly polarized Zeeman signatures can be seen in LSD profiles, at the radial velocity of EK~Dra (at around -20.3~km~s$^{-1}$). 

\begin{figure*}
\epsscale{1.0}
\plotone{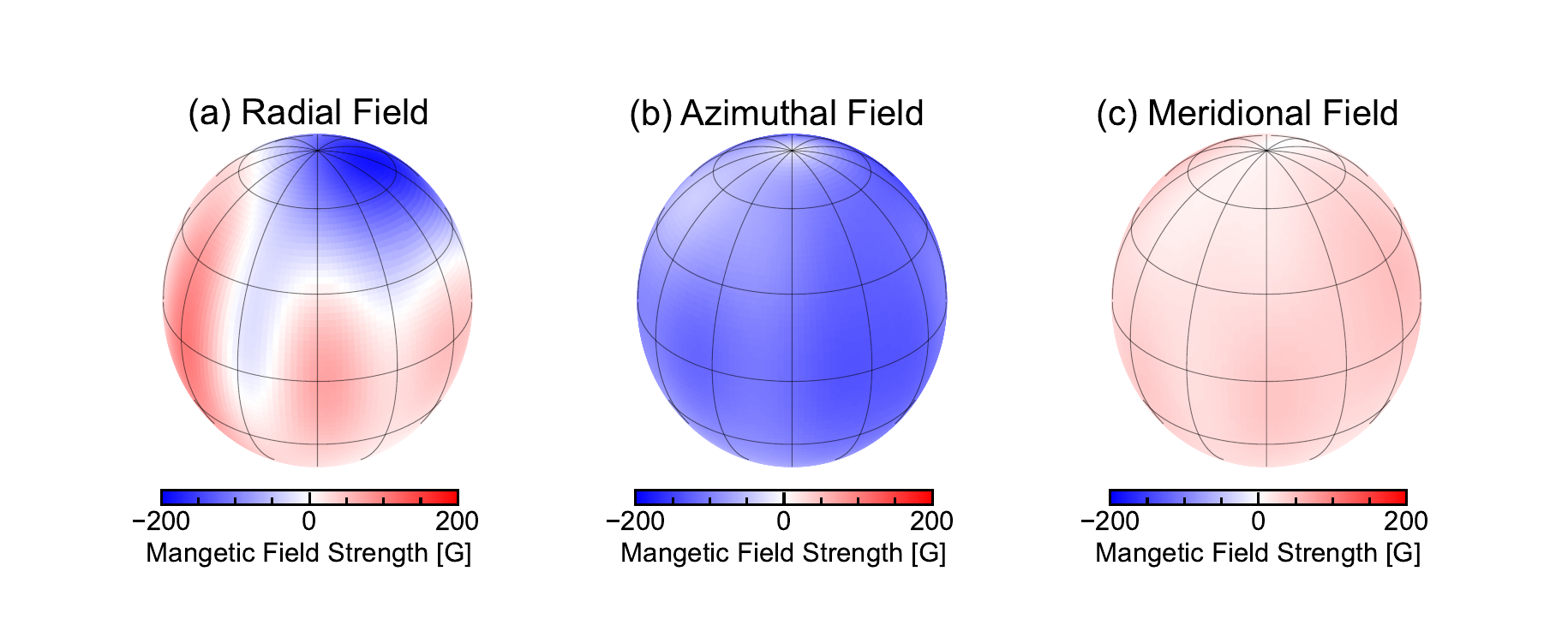}
\caption{
Magnetic field maps on April 10, 2022, for (a) radial, (b) azimuthal, and (c) meridional components.
The maps take into account the stellar inclination angle and depict the hemisphere visible from Earth when the prominence eruption occurred (BJD-2459680.033).
}
\label{fig:magmap}
\end{figure*}

The next step in the modeling consists of using the whole time series of Stokes I and Stokes V LSD profiles to map the surface brightness of EK~Dra (using Stokes I), and its large-scale surface magnetic field (using Stokes V). The Zeeman-Doppler Imaging technique (ZDI hereafter), first described by \citealt{1989A&A...225..456S}, is employed to perform the tomographic inversion. We used here the ZDI code of \cite{2018MNRAS.474.4956F} in which the surface magnetic field is projected onto a spherical harmonics frame, following the formalism of \cite{2006MNRAS.370..629D}. We assumed an inclination angle $i=60^\circ$ and a projected rotational velocity $v \sin i = 16.4$~km~s$^{-1}$, taking our values from the ZDI study of \cite{2017MNRAS.465.2076W}. The rotation phases were calculated assuming a rotation period of $2.766$~d (taken from the same authors) and a reference Julian date equal to 2459661.39. Contrary to \cite{2017MNRAS.465.2076W}, our search for surface differential rotation using the sheared image approach of \cite{2002MNRAS.334..374P} did not deliver any conclusive results using Stokes V spectra, so solid body rotation was assumed in our model, leading to a reduced $\chi^2$ equal to 1.15. 
Figure \ref{fig:magmap} shows the magnetic geometry centered at the rotation phase when the prominence eruption occurred. 
The modeled Stokes V LSD profiles are overplotted on the observed LSD profiles in Figure \ref{fig:lsdV}.

Two main features dominate the complex magnetic geometry reconstructed here. The first one is a strong patch of negative radial field at high latitudes. The second one is a prominent, negative toroidal field component storing 73\% of the surface magnetic energy (the presence of this azimuthal field component is readily visible as a symmetric shape of Stokes V profiles, featuring most of the time a central positive lobe surrounded by two negative lobes). The average unsigned surface magnetic field strength is 120~G, while the maximal unsigned field strength reaches 216~G. The field configuration is mostly axisymmetric, as seen in a majority of stars with a predominant toroidal field component \citep{2015MNRAS.453.4301S}. 
The general field characteristics obtained with our set of observations in 2022 are in overall agreement with the maps reconstructed with data ranging from 2006 to 2012 by \cite{2017MNRAS.465.2076W} (who repeatedly reported a negative toroidal field and a patch of negative radial field close to the pole). The average field value that we obtain here is, however, about 30\% stronger than any average value listed by \cite{2017MNRAS.465.2076W}. 

\begin{figure}
\epsscale{1.0}
\plotone{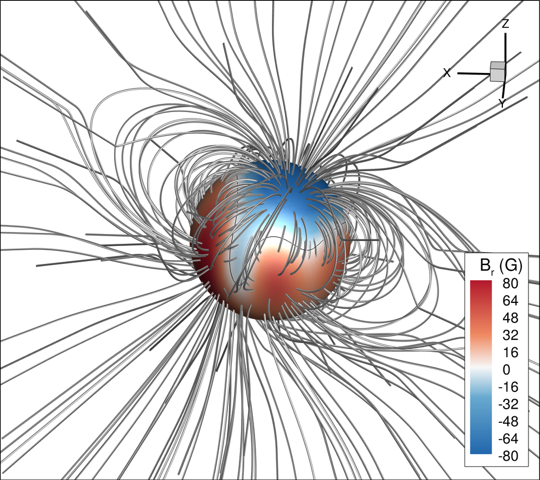}
\caption{Coronal extrapolation of the radial surface magnetic field using a potential field source surface model \citep[e.g.][]{1977SoPh...51..345A}. Here, we assume that the field becomes open and fully radial beyond 3.4 $R_{\rm star}$, mimicking the effects of a stellar wind blowing open the field lines. 
The background surface radial magnetic field is the same as Figure \ref{fig:magmap} and the map depict the hemisphere visible from Earth when the prominence eruption occurred (BJD-2459680.033).
}
\label{fig:potential-field}
\end{figure}

In Figure \ref{fig:potential-field}, coronal extrapolation of the radial surface magnetic field was performed by using a potential field source surface model \citep[e.g.,][]{1977SoPh...51..345A}.
The distance beyond which the field lines become open is a free parameter of our potential field extrapolation. 
This parameter can only be estimated through a stellar wind modelling, which is beyond the scope of the current paper. 
In the solar case, a value of $\sim 2.5~R_{\odot}$ is usually assumed and is supported by observations of the positions of the coronal holes and streamers \citep[e.g.][]{2003SoPh..212..165S,2012ApJ...757...96D}. 
In the case of stars, this value is unknown and likely depends on the stellar wind properties, as well as other stellar properties such as rotation and magnetic field strength \citep[e.g.,][]{2013A&A...557A..67V, 2016ApJ...832..145R}. 
Here, we adopted a value of 3.4 $R_{\rm star}$, as used in other stellar studies \citep[e.g.,][]{2002MNRAS.333..339J}.
therefore, mimicking the effects of a stellar wind blowing open the field lines.
This model provides a simple visualisation of the open and closed coronal magnetic field aiding in the interpretation of the relation between the large-scale field geometry and the surface spot features (see Section \ref{sec:4-1} and \ref{sec:4-2}).

\begin{figure*}
\epsscale{1.0}
\gridline{
\fig{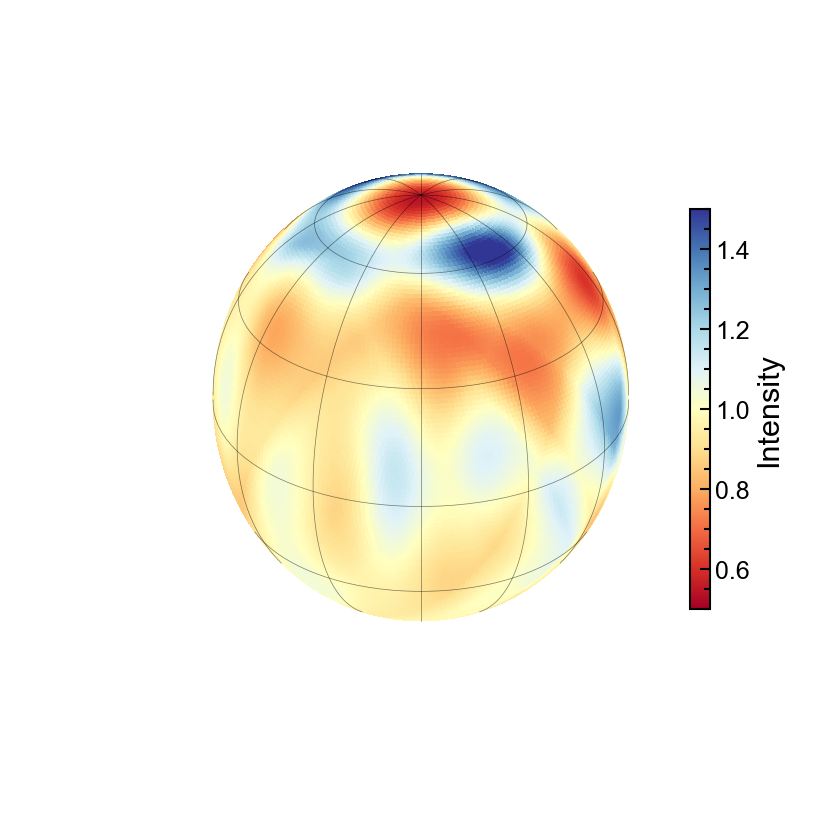}{0.33\textwidth}{\vspace{0mm} (a) Doppler Imaging}
\fig{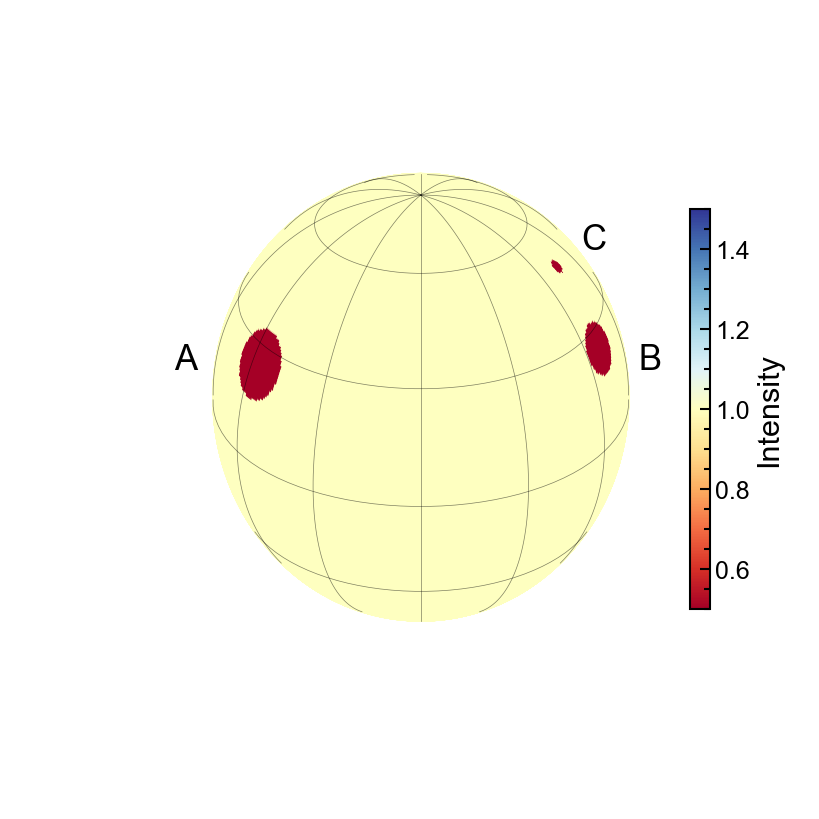}{0.33\textwidth}{\vspace{0mm} (b) TESS Light Curve Modeling}
\fig{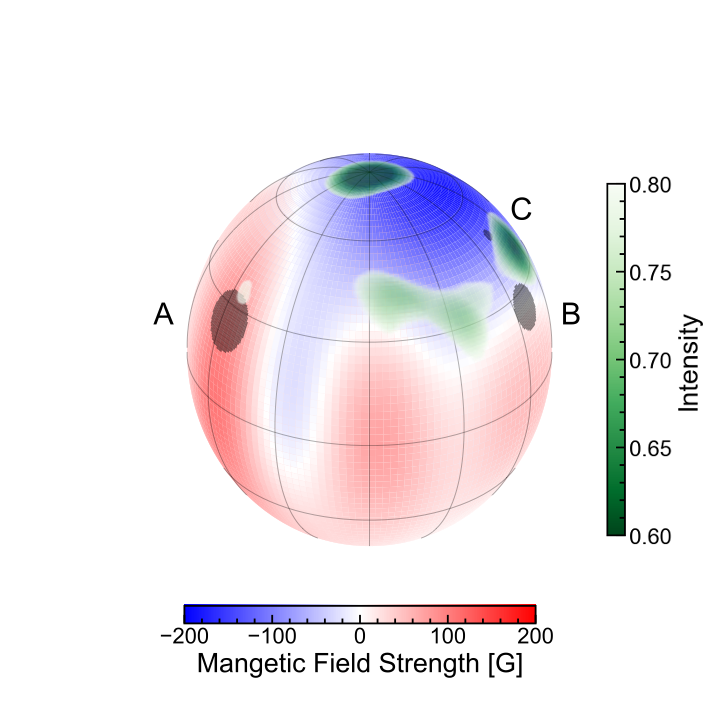}{0.33\textwidth}{\vspace{0mm} (c) ZDI+DI+TESS}
}
\caption{
Starspot maps on April 10, 2022, from (a) Doppler imaging and (b) TESS light curve modeling.
(c) Comparison of the spots from the Doppler imaging (green, right color bar) and those from the TESS light curve modeling (grey circles), plotted on the radial magnetic field (background red and blue colors, the same as Figure \ref{fig:magmap}(a)).
The hemisphere is the same as Figures \ref{fig:magmap} and \ref{fig:potential-field}, a visible hemisphere when the prominence eruption occurred (BJD-2459680.033).
\Add{The values in the intensity map are dimensionless as they are normalized by the quiescent surface intensity of the star.}
}
\label{fig:starspots}
\end{figure*}

Using our set of Stokes I LSD profiles in Figure \ref{fig:lsdI}, we also reconstructed a brightness map of EK~Dra using the Doppler Imaging method \citep[DI hereafter]{1983PASP...95..565V}. The brightness model allowed for both dark and bright spots on the stellar surface, similarly to, e.g., \cite{2020A&A...643A..39C} who used the same code. Figure \ref{fig:starspots}(a) shows the resulting spot distribution centered at the rotation phase when the prominence eruption occurred. 
The polar region is occupied by the darkest spot on the map. This polar feature is coincident, on the magnetic map, with the patch of strong radial field close to the pole, although the latitudinal extension of the magnetic spot is larger, with reaching to a latitude of 30$^\circ$. 
Most bright spots are gathered at a latitude of about 60$^\circ$, while lower latitudes are populated by a mix of bright and dark patches. A second DI model (not shown here) was produced with the inclusion of dark spots only, leading to a surface distribution of dark patches very similar to the one obtained when bright features were allowed as well, at the cost of a 5\% increase of the $\chi^2$. We also searched again for the signature of differential rotation using the Stokes I time series, but failed to identify a unique solution. As for the magnetic map, solid body rotation was therefore assumed here.
Most published brightness maps of EK~Dra show large spots at high latitudes, as reconstructed here (\citealt{1998A&A...330..685S, 2017MNRAS.465.2076W, 2021MNRAS.502.3343S}), with the notable exception of \cite{2018A&A...620A.162J}. However, the high latitude spot is, in most cases, reported to be off-centered with respect to the pole, while the very symmetric configuration reported here is closer to \cite{1998A&A...330..685S} and \cite{2021MNRAS.502.3343S}.

\subsection{Starspot Mapping from TESS Light Curve}\label{section:2-2}


EK Dra (TIC 159613900) was observed by TESS in Sector 50, starting from 26 March to 22 April 2022\footnote{\url{https://archive.stsci.edu/missions/tess/doc/tess_drn/tess_sector_50_drn72_v02.pdf}}. 
In \textsf{Paper \hyperref[2022ApJ...926L...5N]{I}}, the full-frame image (FFI) data obtained with 10 min cadence was processed by using the \textsf{eleanor} package \citep{2019PASP..131i4502F}.
Here we use the ``principal component analysis" (``PCA") flux, which has common systematics between targets on the same camera removed\footnote{Note that in \textsf{Paper \hyperref[2022ApJ...926L...5N]{I}}, we used the combination of raw flux and smoothed PCA flux to extract the small-amplitude flaring light curve because of the relatively large noise in PCA flux, but here we use the unsmoothed PCA flux.}.
The TESS light curve obtained exhibits quasi-periodic rotational modulation with an amplitude of a few percent, which is attributed to surface starspots. 
The light curve of EK Draconis is not entirely periodic, indicating the evolution of starspots and/or different periodicities of each starspot due to surface differential rotation\footnote{We understand that the interpretation of periodic variations in the light curves may not be so straightforward and there are caveats with light-curve fitting \citep[e.g.,][]{2009AIPC.1094..664J,2018ApJ...865..142B}, but here we assume simple starspot modeling.}.

Here we performed a starspot mapping from TESS light curve by using the code implemented by our previous study \citep{2020ApJ...902...73I}. For detailed information on the code, please refer to \cite{2020ApJ...902...73I}.
In the following, we provide an overview of the code, including its assumptions and parameters.
We set certain stellar parameters as fixed, including the contrast of starspots as 0.27 ($T_{\rm spot}$ = 4069 K), which was calculated based on the stellar surface temperature (here we used $T_{\rm eff}$ = 5700 K) using the empirical rule from Equation (4) in \cite{2021ApJ...907...89H}. 
The limb darkening law was assumed based on the empirical rules by \cite{2023A&A...674A..63C}.
The stellar inclination angle $i$ is also fixed as 60$^\circ$ based on previous studies.
The code estimates several parameters using the Markov Chain Monte Carlo (MCMC) method \Add{\citep{2013PASP..125..306F,2017ARA&A..55..213S}}. Despite the vastness of the parameter space, the introduction of the parallel tempering technique \Add{\citep{1996JPSJ...65.1604H,2017ARA&A..55..213S}} allows for a comprehensive coverage of this space within a practical time frame. 
The model for the rotational modulation in TESS light \Add{curves} employs the {\sf macula} model \citep{2012MNRAS.427.2487K}, which assumes circular spots with predefined contrast values, and the emergence/decay of these spots are modeled as a linear emergence/decay in spot area. 
The free parameters include the equatorial rotation period ($P_{\rm eq}$ in days), the degree of differential rotation ($\kappa$), the latitude ($\phi$ in degrees), the initial longitude ($\Lambda$ in degrees), the reference time ($t_{\rm ref}$ in days; the midpoint of the interval during which the spot is at its maximum radius), the maximum radius ($a_{\rm max}$ in degrees), the emergence duration (in days), and the decay duration (in days).

We exclusively utilized data from TESS Sector 50, spanning approximately 27 days. 
We removed large stellar flares from the light curve before the modeling.
EK Dra was also observed in Sectors 48 and 49, but we limited the used data only to Sector 50\Add{, where all of the H$\alpha$ flare data (see \textsf{Paper \hyperref[2022ApJ...926L...5N]{I}}) and most of the DI/ZDI data were obtained}. 
The reason why we limited the period was because extending the timeframe would necessitate the inclusion of more starspots, complicating the resolution of parameter degeneracies \Add{and making it more difficult to find the most likely solution with MCMC. 
To test the impact of using only the limited period on the mapping, we added approximately half of the Sector 49 data to the Sector 50 data and performed the mapping. As a result, the outcomes did not significantly differ. Therefore, the variation in observation time has minimal impact on the results, and we present only the essential data from Sector 50 in this study.} 
In this study, we performed the MCMC sampling with configurations of two and three starspots. Based on the Bayesian criterion ``Evidence", the scenario involving three starspots was adopted \citep{2020ApJ...902...73I,2023ApJ...948...64I}.

\begin{figure*}
\epsscale{1.0}
\plotone{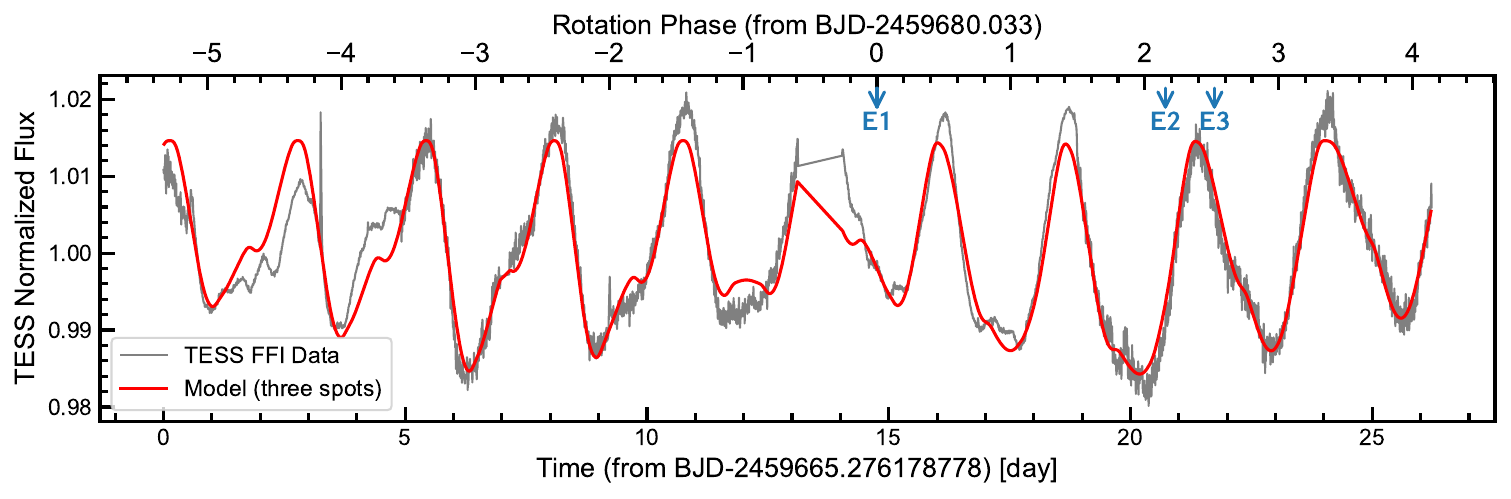}
\caption{Results from TESS light curve modeling. The gray line represents the observed data from TESS Full Frame Images (FFI) in 2022 (Sector 50), and the red line shows the modeled light curve, which includes three spots.
}
\label{fig:lc-model}
\end{figure*}

\begin{deluxetable}{cccccccc}
\label{tab:spotparams}
\tablecaption{Spot parameters on April 10, 2022 from TESS light curve inversion.}
\tablewidth{0pt}
\tablehead{
\colhead{} & \colhead{Spot A} & \colhead{Spot B} & \colhead{Spod C}   
}
\startdata
Length Scale [10$^{10}$ cm] &  1.02$^{+ 0.08}_{-0.07}$ & 0.79$^{+0.05}_{-0.07}$ & 0.22$^{+0.02}_{-0.02}$ \\
(Length Scale [R$_{\rm star}$]) & 0.1565$^{+0.0001}_{-0.0001}$ & 0.1192$^{+ 0.0001}_{- 0.0001}$ & 0.034$^{+0.0001 }_{-0.0001}$ \\
Latitude [deg] & 25.06$^{+0.08}_{-0.02}$ & 25.05$^{+0.09}_{-0.02}$ & 48.13$^{+0.06}_{-0.07}$ \\
Longitude [deg]$^\dagger$ & -56.84$^{+0.01}_{-0.06}$  & 67.15$^{+0.06}_{-0.01}$  & 72.71$^{+0.06}_{-0.06}$  \\
\enddata
\tablecomments{$^\dagger$Longitude is defined by setting the prime meridian, which is the meridian at 0 degrees, as the reference point when viewed from the Earth.}
\end{deluxetable}

Figure \ref{fig:lc-model} presents a comparison between the TESS light curve and the optimum light curve. 
While some residuals remain, the overall characteristics of the light curve are well reproduced. 
It has been reported that increasing the number of starspots could fill in these residuals with smaller spots \citep{2023ApJ...948...64I}, \Add{but} even our three-spot model adequately reproduces the general features of the light curve. 
Table \ref{tab:spotparams} summarizes the estimated spot length scale (radius), latitude, and longitude, at the time of the prominence eruption.
Other important parameters, such as differential rotation and emergence/decay rates, will be presented in our next paper (Ikuta et al. in prep.)
Figure \ref{fig:starspots} illustrates the configuration of starspots at the time of the prominence eruption.
We can see that there are three mid-latitude spots (labeled A, B, and C in Table \ref{tab:spotparams}), all of which are in the visible hemisphere at this time.

In the following, we compare this result with the DI/ZDI map in Section \ref{sec:4-1} and compare them with prominence eruption in Section \ref{sec:4-2}.


\section{Observationally Constrained Model of a Prominence Eruption}\label{sec:3}

\subsection{Observation Summary}\label{sec:3-1}

The observational characteristics of the superflare event on the young solar-type star EK Dra are summarized below: 
The superflare (refered to as ``E1" in \textsf{Paper \hyperref[2022ApJ...926L...5N]{I}}) occurred on April 10, 2022 during a multiwavelength observation campaign spanning from April 10 to April 21, 2022. 
The TESS white-light flare exhibited an energy magnitude of \(1.5 \times 10^{33}\) erg, classifying it as a superflare.
During the superflare, significant increases in H$\alpha$ emission were observed. 
The event was characterized by a strong blueshifted emission in the H$\alpha$ line, indicating a rapid outward movement of material at LOS speeds reaching up to 690 km s$^{-1}$. 
This blueshifted emission was interpreted as the first detection of a prominence eruption from a young solar-type star. 

\begin{figure*}
\epsscale{0.7}
\gridline{
\fig{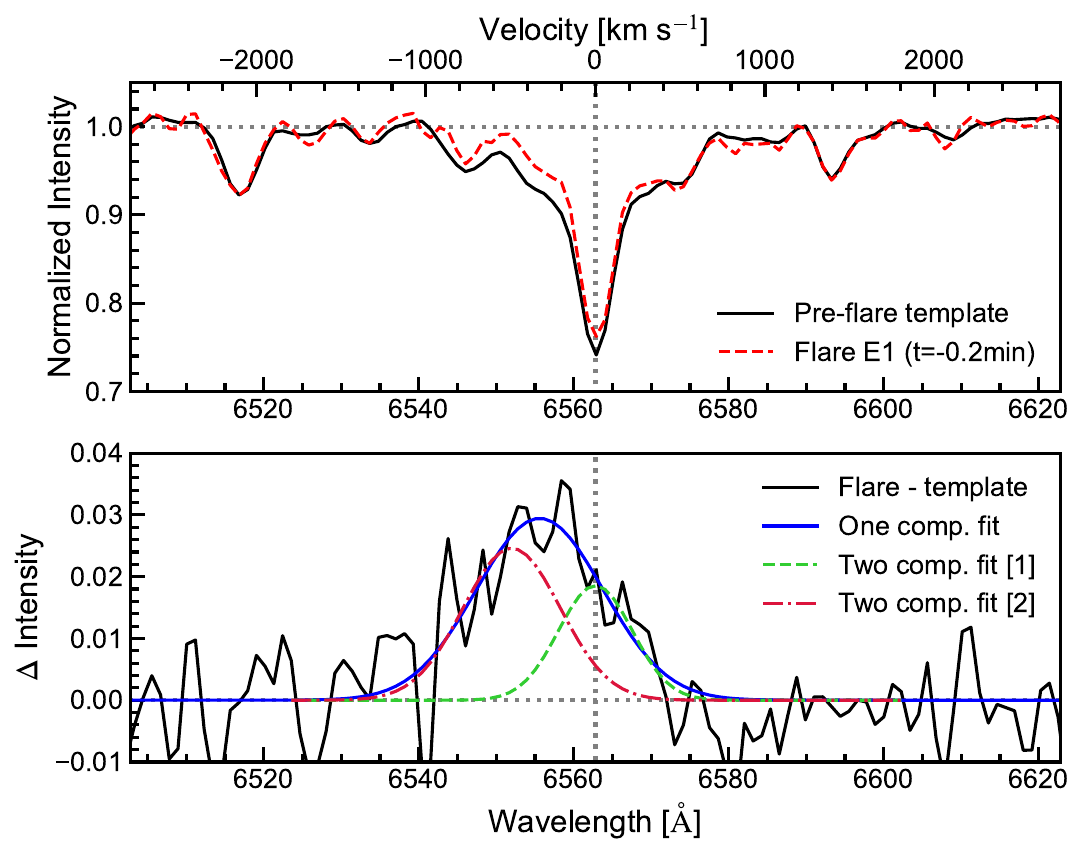}{0.5\textwidth}{\vspace{0mm} (a)}
\fig{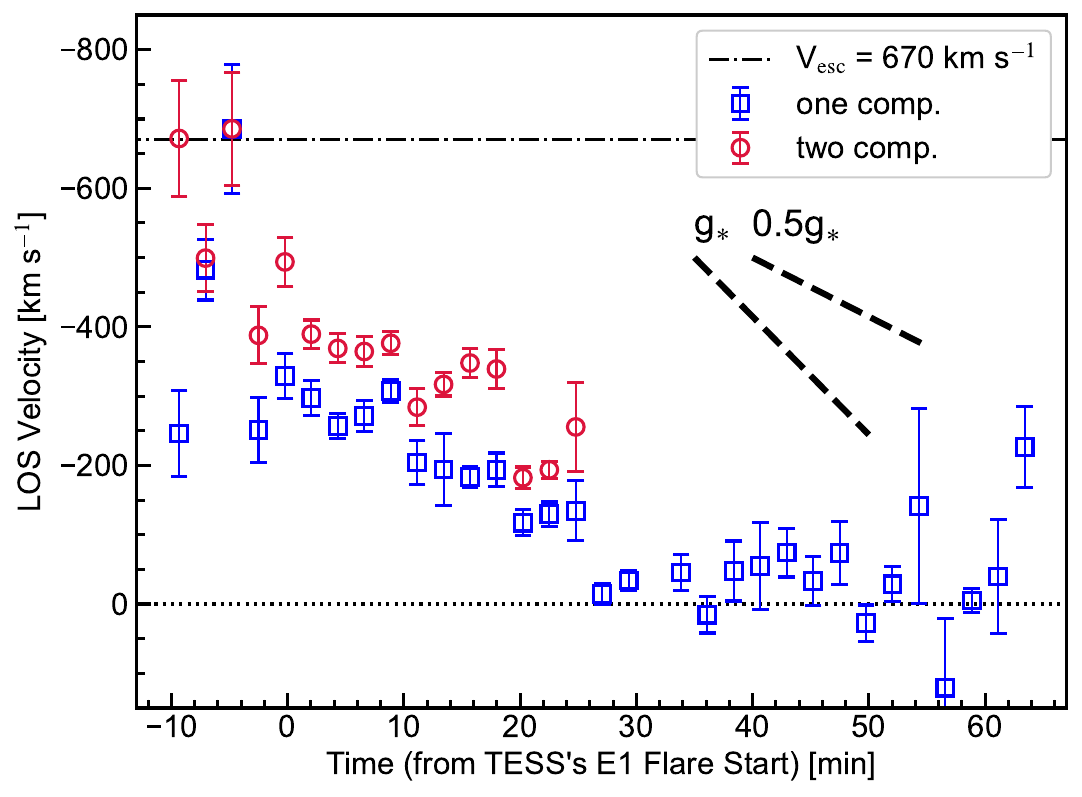}{0.5\textwidth}{\vspace{0mm} (b)}
}
\caption{An example of (a) the blueshifted H$\alpha$ spectra and (b) the time evolution of the velocity taken from \textsf{Paper \hyperref[2022ApJ...926L...5N]{I}}. }
\label{fig:halpha-obs-summary}
\end{figure*}

Figure \ref{fig:halpha-obs-summary} shows an example of the observed H$\alpha$ line profiles and the temporal changes in velocity (taken from \textsf{Paper \hyperref[2022ApJ...926L...5N]{I}}\footnote{In Figure 6(c) and 7(c) of \textsf{Paper \hyperref[2022ApJ...926L...5N]{I}}, the integrated time is described as $\pm$5 min, but correctly, it is $\pm$2.5 min. Here, we correct this typo.}).
In \textsf{Paper \hyperref[2022ApJ...926L...5N]{I}}, as in Figure \ref{fig:halpha-obs-summary} (b), the blueshifted H$\alpha$ profiles were analyzed using single Gaussian fitting (hereafter ``one component" fitting) or two Gaussian fitting (hereafter ``two component" fitting) methods for the velocity estimates, depending on the different interpretation of the central component.
For two-component fitting, it is assumed that emissions near the line center originate from flare ribbon emissions, and additional blueshifted emissions are superimposed on the profile. Based on this assumption, the line center component is presumed to have a velocity of zero, making the velocity of the blueshifted component the only free parameter.
As a result of these fittings, Figure \ref{fig:halpha-obs-summary} (b) shows two types of velocity evolution. 
Both methods estimate the maximum LOS velocity\footnote{In \textsf{Paper \hyperref[2022ApJ...926L...5N]{I}}, there was a typo that the error bars were described as 690${\pm (92-93)}$ km s$^{-1}$ in some parts. Here, we correct this typo.} of 690${\pm (82-93)}$ km s$^{-1}$, which is notable as it approaches the stellar escape velocity $\sim$670 km s$^{-1}$. 
The decrease in velocity over time was comparable to the stellar gravity, suggesting that the prominence material's dynamics were influenced by the star's gravitational field.
However, after the blueshift lasting for $\sim$20 min, the emission component approached $\sim$0 km s$^{-1}$, and eventually disappeared without showing any redshift.
The reasons behind the disappearance of the emission at a velocity of zero remained unclear, posing questions about the dynamics.

\begin{deluxetable*}{lccccccc}
\label{tab:length-scale}
\tablecaption{Summary of length scale and area of the superflares, prominence/filament eruptions, and starspots on EK Dra estimated in \textsf{Paper \hyperref[2022ApJ...926L...5N]{I}}.}
\tablewidth{0pt}
\tablehead{
\colhead{Date (ID)} & \colhead{BJD} & \colhead{Blueshift} & \colhead{Flare Loop} & \colhead{Prominence} & \colhead{Spot} & \colhead{Rot. Phase} & \colhead{LC Profile}  \\
\colhead{} & \colhead{} & \colhead{} & \colhead{[10$^{10}$ cm]} & \colhead{[10$^{10}$ cm]} & \colhead{[10$^{10}$ cm]} & [0-1] & \colhead{} \\
}
\startdata
2022 Apr 10 (E1) & 2459680.033 & Yes & 0.73  & 8.3-300  & 2.1 (0.22-1.02)$^\ddagger$ & 0.00 & Decline-Loc.min \\
2022 Apr 16 (E2) & 2459685.998 & Yes & 1.16  & 1.4-51  & 2.4 & 0.15 &  Rise \\
2022 Apr 17 (E3) & 2459687.012 & No & 0.67  & --  & 2.4 & 0.51 &  Decline \\
2020 Apr 05 (E4) & 2458945.241 & Yes & 0.48  & 1.4-40$^{\ast}$ & 1.9$^{\ss}$ & 0.73$^\dagger$ &  Loc.max \\
2020 Mar 14 (E5) & 2458923.004 & No & 1.50  & -- & 2.3$^{\ss}$ & 0.70$^\dagger$ &  Decline \\
\enddata
\tablecomments{
The table data is taken from \textsf{Paper \hyperref[2022ApJ...926L...5N]{I}}.
``Rotation Phase" is the timing when the superflares occur relative to stellar rotational phase. Here, rotational phase 0 is defined as 2459680.033 and rotational period $P_{\rm rot}$ is defined as 2.77 days (i.e, phase = ((BJD - 2459680.033) mod $P_{\rm rot}$)/$P_{\rm rot}$). 
$^\dagger$Given that the observations are conducted two years apart from the reference time, these values can be meaningless considering the change in spots and differential rotation.
``LC Profile" is the light curve profile when the superflare occur. ``Loc.min" and ``Loc.max" is the local minimum and local maximum of the light curve, respectively, and ``Rise" and  ``Decline" is the rising and decline phase of the light curve, respectively.
As a reference, stellar radius is 0.94 times solar radius $\sim$6.55$\times 10^{10}$ cm, and stellar disk area is $\sim$13.5$\times 10^{21}$ cm$^{2}$.
$^{\ast}$There was a mistake in the estimation of upper limit of length scale in \textsf{Paper \hyperref[2022ApJ...926L...5N]{I}}, so we revised here.
$^\ddagger$The spot length scale from the TESS light curve modeling, as described in Table \ref{tab:spotparams}.
$^{\ss}$The spot length scales of E4-E5 events are recalculated to follow the calculation of E1-E3 events, i.e., the local maximum minus local minimum of the light curve over one rotation containing a flare. However, since the E4 event includes a flare around the local maximum, the maximum value was defined as the top 98\% of brightness within one rotation.
}
\end{deluxetable*}

These observational \Add{signatures indicate} the following strong constraints on the spatial and dynamic properties of the eruption:
\begin{enumerate}
    \item First, the presence of an H$\alpha$ emission profile without any absorption signatures indicates that the prominence was consistently positioned above the stellar limb throughout the eruptive event. 
    \item Second, the detection of a white-light flare suggests that the flare’s footpoints were not entirely obscured by the stellar disk. 
    \item Additionally, the observed deceleration of the prominence, which aligns closely with the star's surface gravity, implies that the eruption could be oriented nearly along the line of sight. 
\end{enumerate}
Based on the above constraints, in the following sections we will estimate the possible dynamics of prominence eruption, such as the event's location, direction, and evolution, by using two different types of simplified models.

\subsection{Simple One-Dimensional Free-fall Model}\label{sec:3-2}

As a first approach, we consider a simple model where prominence center-of-mass is ejected in a one-dimensional direction and fall solely under the influence of stellar gravity (hereafter referred to as the simple ``one-dimensional free-fall" model). 
It should be noted that this is not a complete ``free-fall" scenario, as it does not take into account the gravitational forces perpendicular to the path of motion. 
In the case of the Sun, the ejection of prominences is influenced by the configuration of magnetic field lines. Some erupted prominences land on another point of the solar surface but others \Add{return} near their original footpoints along magnetic lines. 
This case is poorly understood for stellar eruptive events, therefore, to avoid complexities arising from the lack of observational constraints on the stars, we
have adopted this simplified approach.
Our study is the first attempt of this kind of dynamical modelling among stellar flare/prominence eruptions observations. 
This is because of our well-defined \Add{characteristics} of the stellar prominence eruptions as summarised in Section \ref{sec:3-1}. 

\begin{figure}
\epsscale{1.0}
\plotone{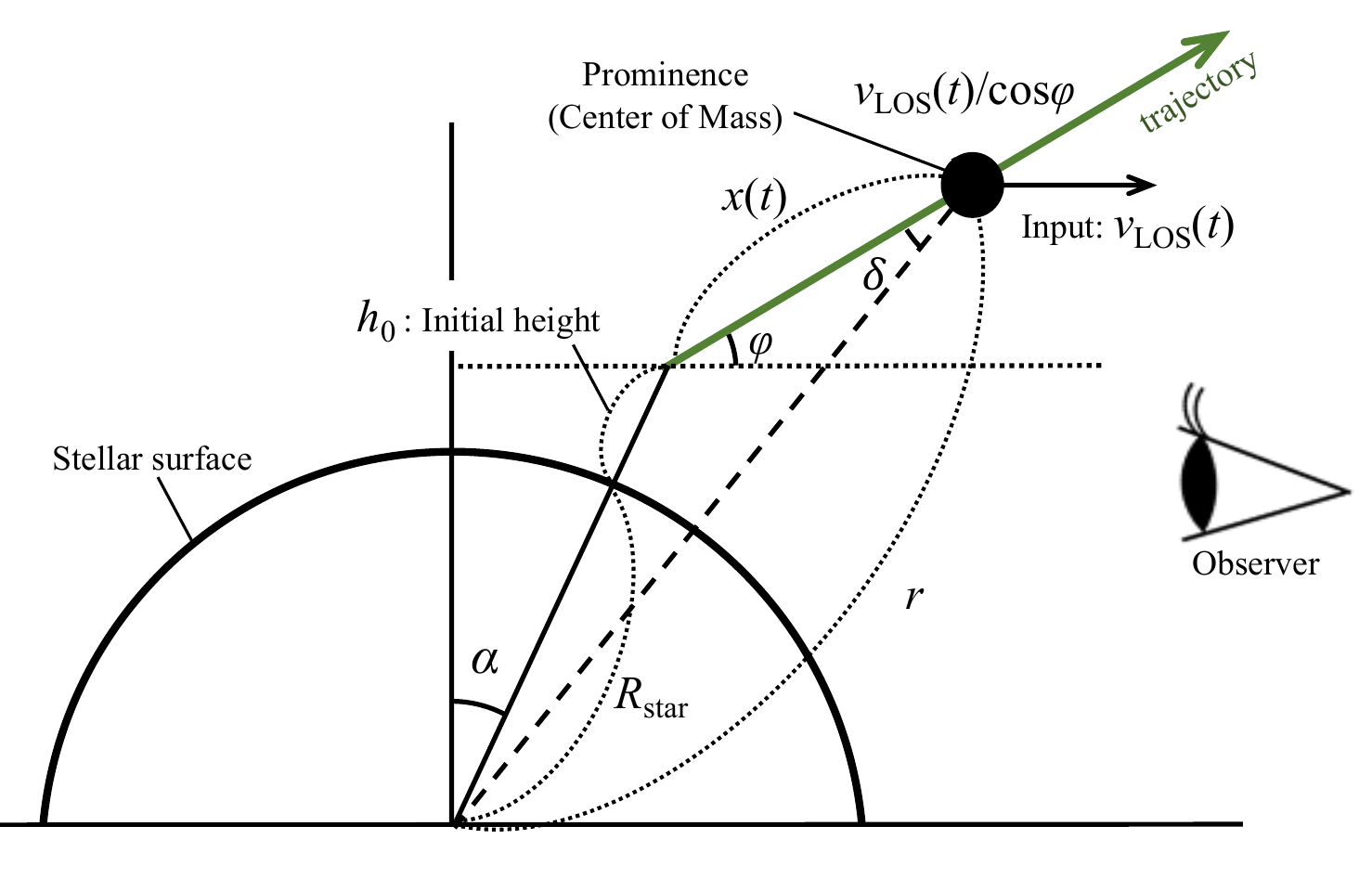}
\caption{Schematic picture of the simple ``one-dimensional free-fall" model. 
}
\label{fig:schematic-model}
\end{figure}

Observations provide a well-defined time variation of LOS velocity of prominence center-of-mass $v_{\rm LOS}(t)$. 
To reproduce $v_{\rm LOS}(t)$ using the schematic model described in Figure \ref{fig:schematic-model}, which satisfies conditions 1 to 3 mentioned above, we estimate the optimal values of the following four parameters using \Add{an} MCMC:
\begin{itemize} 
    \item $h_0$: Initial height of the prominence at $t = t_0$, where $h_0 > 0$,
    \item $\phi$: The angle of prominence eruption direction against LOS direction of observer, where $0< \phi < \pi/2$,
    \item $\alpha$: The angle of initial prominence direction against stellar limb direction, where $0< \alpha < \pi/2$
    \item $v_0$: Initial LOS velocity ($t=t_0$)
\end{itemize}
Here, in the case of the one-component fit for H$\alpha$ line profiles (see Section \ref{sec:3-1}), the used time range is $t_0 = -0.23$ min $< t < 52.03$ min for optimizing, and in the case of the two-component fit, the range is $t_0 = -0.23 $ min $< t < 24.79$ min. 
As for the one-component fit data, before time $-0.23$ and after 52.03 min, the signal is weak, resulting in a large error in velocity, so we did not include these periods.
As for the two-component fit data, after 24.79 min, the residual asymmetric components were not obtained.

In the schematic picture in Figure \ref{fig:schematic-model}, the distance $r(t)$ from the center of the star and the angle $\delta (t)$ between the radial direction and the direction of the eruption at a given time $t$ and the distance traveled along a straight line from the initial point $x(=x(t))$ (see Figure \ref{fig:schematic-model}) can be expressed as follows:
\begin{eqnarray}
    r(x) &=& \left( x^2 + (h_0 +R_{\rm star})^2 + 2x(h_0 +R_{\rm star})\sin(\phi + \alpha) \right)^{\frac{1}{2}} \label{eq:r} \\
    \cos \delta (x) &=& \frac{x^2 + r(x)^2 - (h_0+R_{\rm star})^2}{2 x r(x)} \label{eq:delta}
    \end{eqnarray}
The equation to be solved is,
\begin{eqnarray}
    \frac{d^2x}{dt^2} &=& - \frac{GM}{r(x)^2}\cos\delta (x) \label{eq:EOM}
\end{eqnarray}
Here, $G$ is the gravitational constant, ($6.67\times 10^{-8}$ cm$^3$ s$^{-2}$ g$^{-1}$), $M$ is the stellar mass (0.95 $M_\odot$), and $R_{\rm star}$ is the stellar radius (0.94 $R_\odot$). 
Equation \ref{eq:EOM} can be solved under a given initial condition of $h_0$, $\phi$, and $\alpha$ by using the fourth-order Runge-Kutta method.
We performed parameter estimations by using the adaptive MCMC methods \Add{\citep{2013PASP..125..306F,ArakiPT2013}}. 
The number of chains is 10000 and burn-in sample size is 2000\footnote{\Add{Burn-in sample size refers to the initial samples in an MCMC simulation that are discarded to minimize the influence of the starting point. This allows the chain to converge to the target distribution. The acceptance ratio is the fraction of accepted steps, indicating the chain's efficiency. A balanced ratio (often $\sim$20-30\%) ensures effective exploration of the parameter space.}}.
The acceptance ratio of the MCMC chain is approximately 0.25.
The proposal distribution is tuned during the first burn-in 2000 chain with the adaptive methods.

Boundary conditions are set as
\begin{itemize}
    \item $ (h_0 + R_{\rm star}) \cos \alpha > R_{\rm star}$: The prominence should be outside the stellar limb from the observer's view,
    \item $0< h_0 < 2 R_{\rm star}$, $0 < \phi < \pi/2$, and $0 < \alpha < \pi/2$.
\end{itemize}
In \textsf{Paper \hyperref[2022ApJ...926L...5N]{I}} we learned that the initial deceleration ($0.34\pm0.15$ km s$^{-2}$) was almost entirely surface gravity ($0.30\pm0.05$ km s$^{-2}$). 
\Add{This indicates that the initial height of the eruption is relatively close to the stellar surface. If we assume the initial height as $h = 2R_\mathrm{star}$, the gravitational deceleration becomes one-fourth, making it unlikely that this is the initial position.}
So we set the boundary condition for the initial height at two stellar radii, but changing this condition did not significantly affect the results.

\begin{figure*}
\epsscale{0.7}
\gridline{
\fig{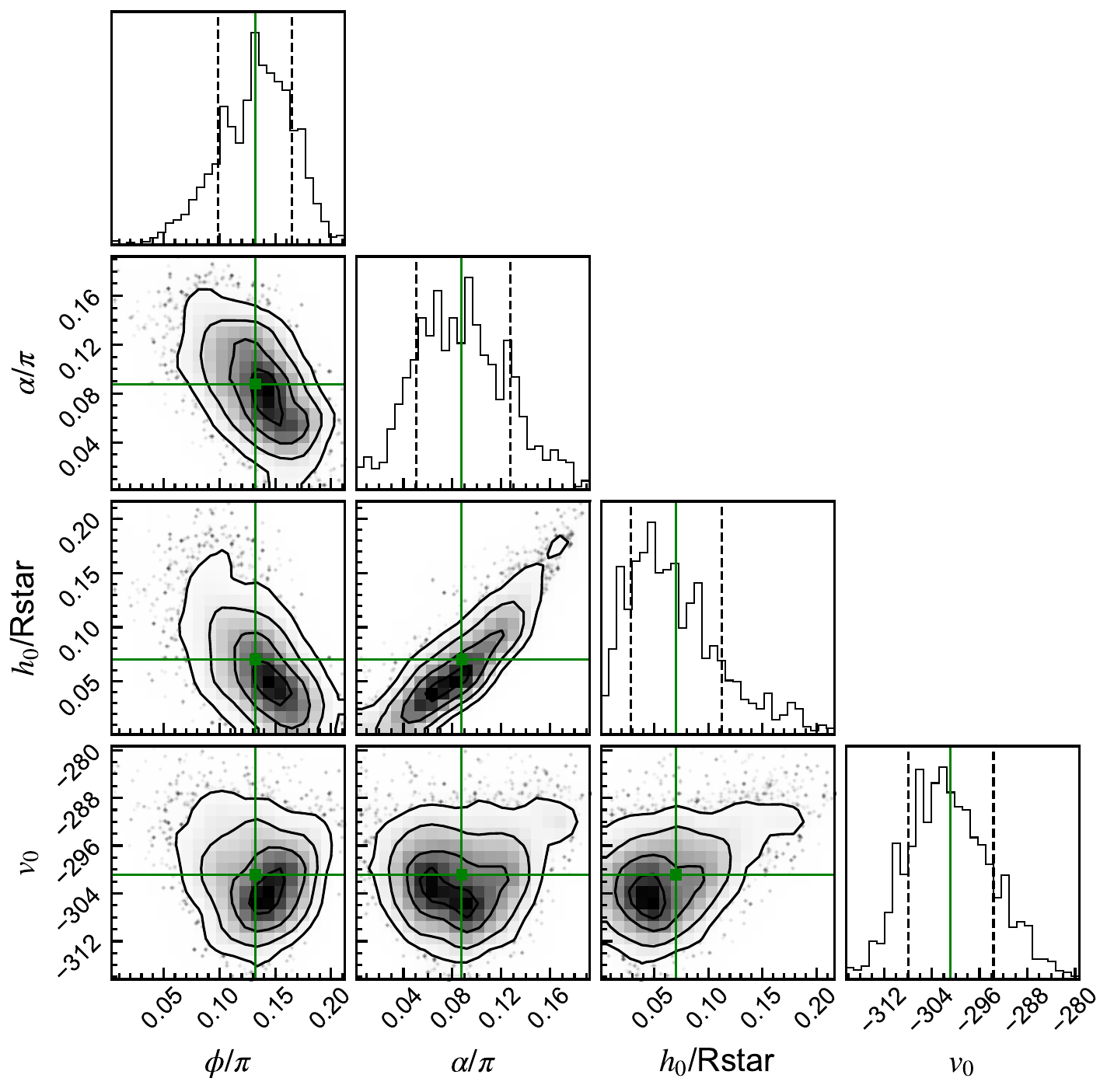}{0.5\textwidth}{\vspace{0mm} (a)}
\fig{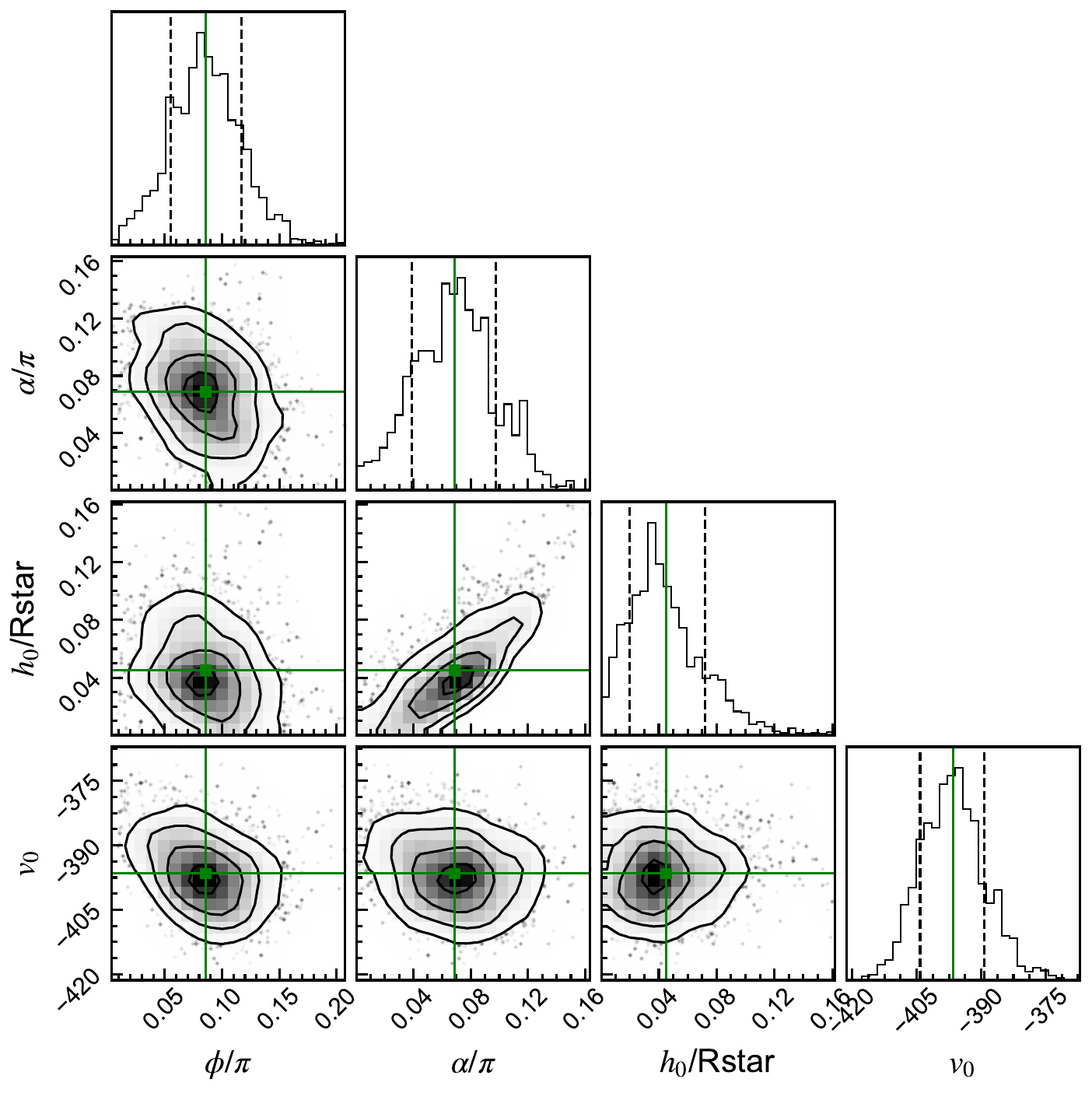}{0.5\textwidth}{\vspace{0mm} (b)}
}
\caption{Correlations between posterior samples for each parameter for ``one-dimensional free-fall" model (Section \ref{sec:3-2}). 
\Add{Panels (a) and (b) show the MCMC sampling results for the velocity evolution obtained by the one- and two-component fitting of the H$\alpha$ line, respectively. 
The green line represents the central value obtained by fitting the histogram to a normal distribution using \textsf{scipy.stats.norm.fit}. The black dashed lines indicate the 16\% and 84\% regions of the distribution.
These boundaries correspond to the 68\% confidence interval, commonly used to represent parameter uncertainty in MCMC analysis.}
The figure is plotted with \textsf{corner} in \textsf{python}.
}
\label{fig:mcmc}
\end{figure*}

\begin{figure*}
\epsscale{0.7}
\gridline{
\fig{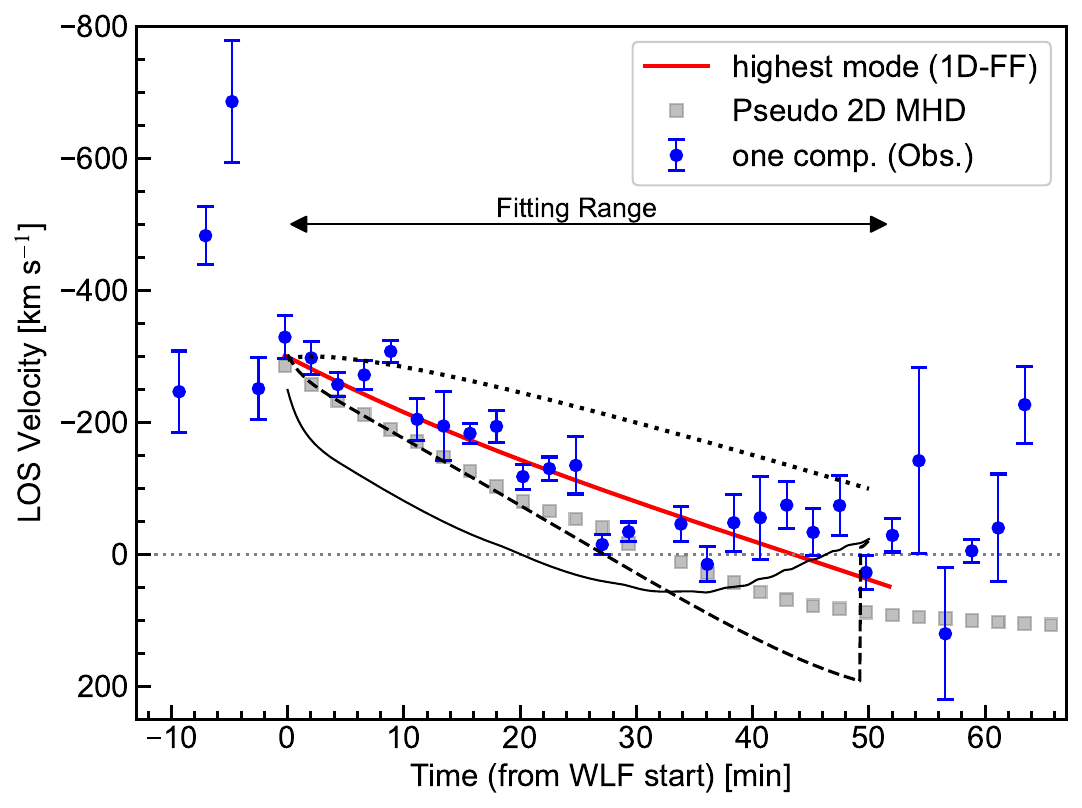}{0.5\textwidth}{\vspace{0mm} (a)}
\fig{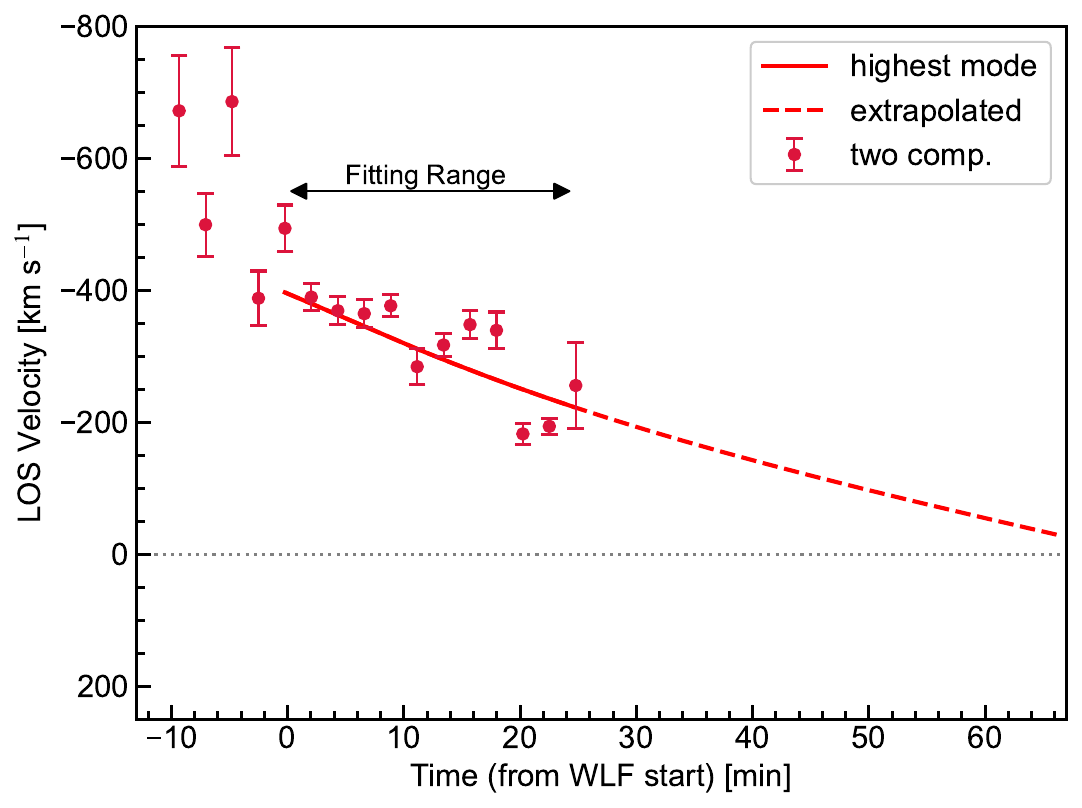}{0.5\textwidth}{\vspace{0mm} (b)}
}
\gridline{
\fig{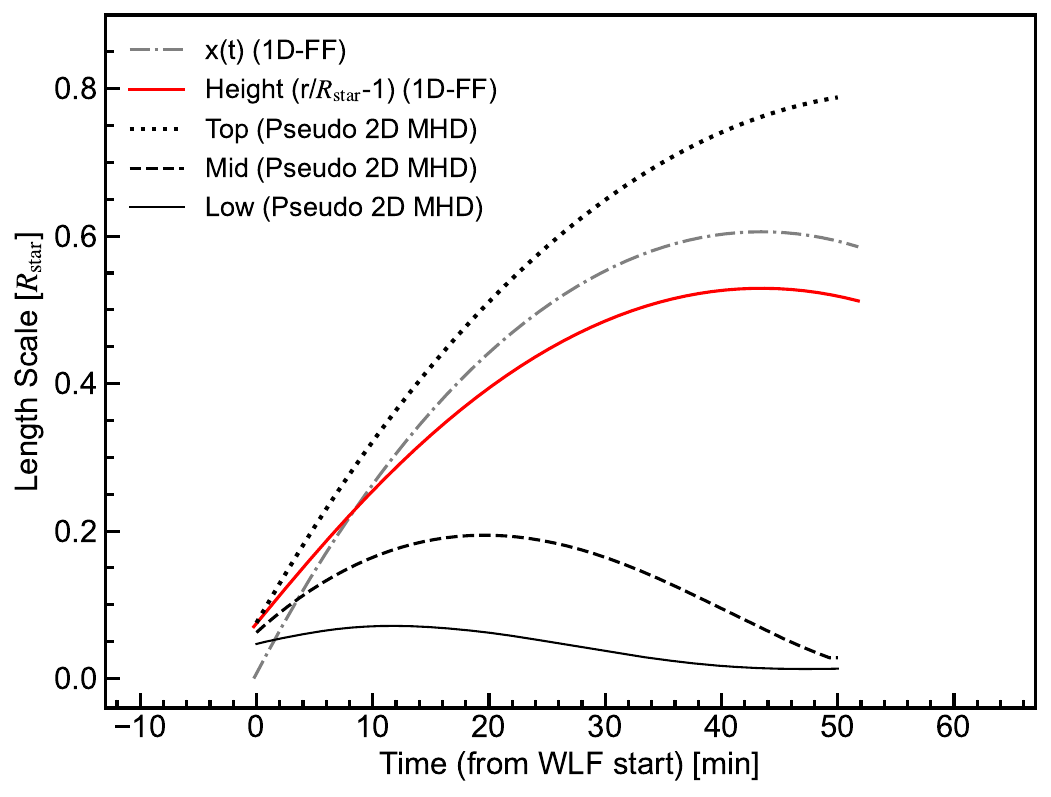}{0.5\textwidth}{\vspace{0mm} (c)}
\fig{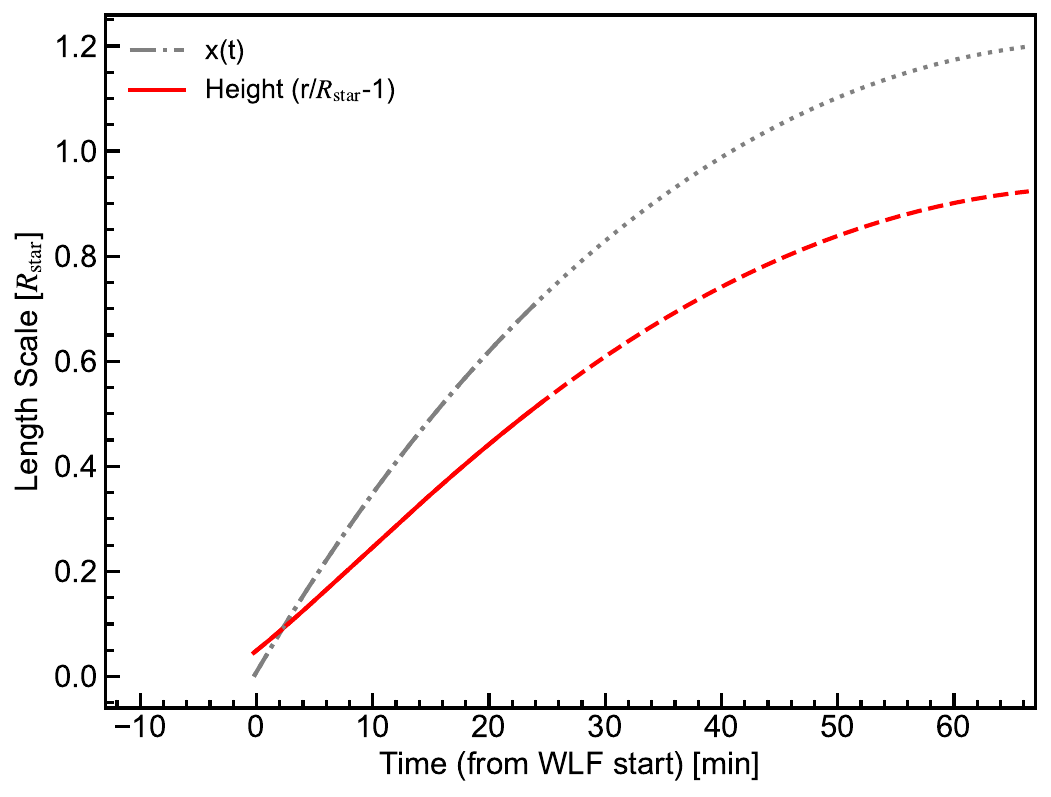}{0.5\textwidth}{\vspace{0mm} (d)}
}
\caption{The velocity and length scale evolution with the estimated parameters from one-dimensional free-fall model (``1D-FF"). 
The \Add{panels} (a) and (b) represent the time evolution of velocity for the one-component and two-component fit for H$\alpha$ line profiles, respectively. 
The modeled time evolution with the most-likely parameters are indicated with \Add{red} solid line and its extrapolation with \Add{red} dashed line. 
The time range used for MCMC sampling is indicated with arrows.
The \Add{panels} (c) and (d) represent the time evolution of the $x(t)$ (grey dash-dotted line) and the prominence height (red solid line) for the one-component and two-component fit for H$\alpha$ line profiles, respectively.
For a reference, panels (a) and (c) depict the components at three locations within a one dimensional hydrodynamic simulation of the flow along the expanding magnetic loop (``Pseudo 2D MHD" model; \citealt{2024ApJ...963...50I} and Ikuta et al. in prep.) from Figure \ref{fig:ikuta-model} using black lines (the line styles are consistent with those in Figure \ref{fig:ikuta-model}). Additionally, in panel (a), the center positions of the Gaussian fits to the pseudo H$\alpha$ from Figure 11(c) are indicated by gray squares.
}
\label{fig:mcmc-velocity}
\end{figure*}

Then we performed two different MCMC runs for the velocity evolution which are 
obtained by \Add{one- and two-component fitting} of the blueshifted H$\alpha$ spectral line profile. 
Figure \ref{fig:mcmc} (a) and (b) show the results of the sampling for the \Add{one- and two-component fitting cases}, respectively.
The estimated parameters are summarized in Table \ref{tab:mcmc-result}.
Figure \ref{fig:mcmc-velocity} compares the observed velocity variations with the modeled one based on the estimated optimal parameters. 
\Add{The figure confirms that the model curve corresponds well with the observed data points.}
From these estimation results, in both cases, a generally similar picture of the eruption is presented: 
\begin{enumerate}
    \item the eruption originates near the stellar surface (0.045--0.07 $R_{\rm star}$),
    \item it erupts in a direction approximately 15--24 degrees from the line of sight, which is relatively close to the line of sight, and
    \item it erupts from a point close to the limb, at an angle of 12--16 degrees from the limb.
\end{enumerate}

\begin{deluxetable*}{lccccccccc}
\label{tab:mcmc-result}
\tablecaption{Deduced parameters for kinematics of eruptive prominence in the schematic model of Figure \ref{fig:schematic-model}.} 
\tablewidth{0pt}
\tablehead{
\colhead{Parameters} & \colhead{Input} & \colhead{Prior dist.} & \colhead{Proposal dist.} & \multicolumn{2}{c}{Estimates from Posterior} \\
\colhead{} & \colhead{} & \colhead{} & \colhead{} & \colhead{(A) One comp. } & \colhead{(B) Two comp.}
}
\startdata
Initial Height $h_0$ [$R_{\rm star}$]  & 0.1$R_{\rm star}$ & ${\cal U} (0,2R_{\rm star})$ & ${\cal N}(\mu,\sigma^2)$ & 0.070$^{+0.042}_{-0.042}$ &  0.045$^{+0.027}_{-0.025}$ \\
LOS angle $\phi$  & 0.1$\pi$ & ${\cal U} (0,\pi/2)$ & ${\cal N}(\mu,\sigma^2)$ & 0.132$^{+0.033}_{-0.033}\pi$  ( 24$^{+6}_{-6}$ deg ) & 0.086$^{+0.031}_{-0.030}\pi$  ( 15$^{+6}_{-5}$ deg ) \\
Angle $\alpha$  & 0.1$\pi$ & ${\cal U} (0,\pi/2)$ & ${\cal N}(\mu,\sigma^2)$ & 0.088$^{+0.040}_{-0.037}\pi$  ( 16$^{+7}_{-7}$ deg ) & 0.069$^{+0.029}_{-0.030}\pi$  ( 12$^{+5}_{-5}$ deg ) \\
LOS Velocity $v_0$ [km s$^{-1}$] & $v_{\rm in}$ & ${\cal U} (v_{\rm min},v_{\rm max})$ & ${\cal N}(\mu,\sigma^2)$ & --301$^{+7}_{-7}$ & --397$^{+7}_{-8}$
\enddata
\tablecomments{
${\cal U}(a,b)=1/(b-a)$ is Uniform distribution. 
$v_{\rm min}=-378$ km s$^{-1}$ and $v_{\rm max}=-278$ km s$^{-1}$ for (A) one component fit model, while $v_{\rm min}=-500$ km s$^{-1}$ and $v_{\rm max}=-300$ for (B) two component fit model.
${\cal N}(\mu,\sigma^2)$ is normal distribution. 
$\sigma$ is different for each parameter.
We defined the center of a posterior distribution fit with a normal distribution as the most likely value. The error range is determined by the region from the 16th to the 84th percentile of the posterior.
}
\end{deluxetable*}

Of course, in reality, the direction of the ejection may not be linear, and the prominence likely has a spread-out structure relative to its center of mass, so it is important to keep in mind that this model is a highly simplified representation. 
However, there have been no studies that could infer information such as the direction of stellar eruptions to this extent, and this work could serve as a basis for more sophisticated model calculations. 
In the next section, we attempt to reproduce the observations using a more refined one-dimensional (1D) hydrodynamic (HD) simulation of the flow along the expanding magnetic loop (pseudo two-dimensional magnetohydrodynamics (2D MHD) model), based on the constrained information obtained here.

\subsection{1D HD Simulation of the Plasma Flow along an Expanding Magnetic Loop (Pseudo 2D MHD Model)}\label{sec:3-3}

Here we present the results of the one-dimensional (1D) hydrodynamic (HD) simulation of the plasma flow along an expanding magnetic loop (a pseudo two-dimensional magnetohydrodynamics (2D MHD) model), emulating the dynamics of a prominence eruption, by using the methodology employed by \cite{2024ApJ...963...50I}.
This is also a simple but more sophisticated model than that described in Section \ref{sec:3-2}, which is based on a solar prominence eruption model.
The aim of this simulation is to investigate whether the observed dynamic spectrum of H$\alpha$, i.e. the velocity change, can be explained by a simple hydrodynamic model, whose initial conditions are constrained on the basis of the results in Section \ref{sec:3-2}.

\begin{figure*}
\epsscale{1.0}
\plotone{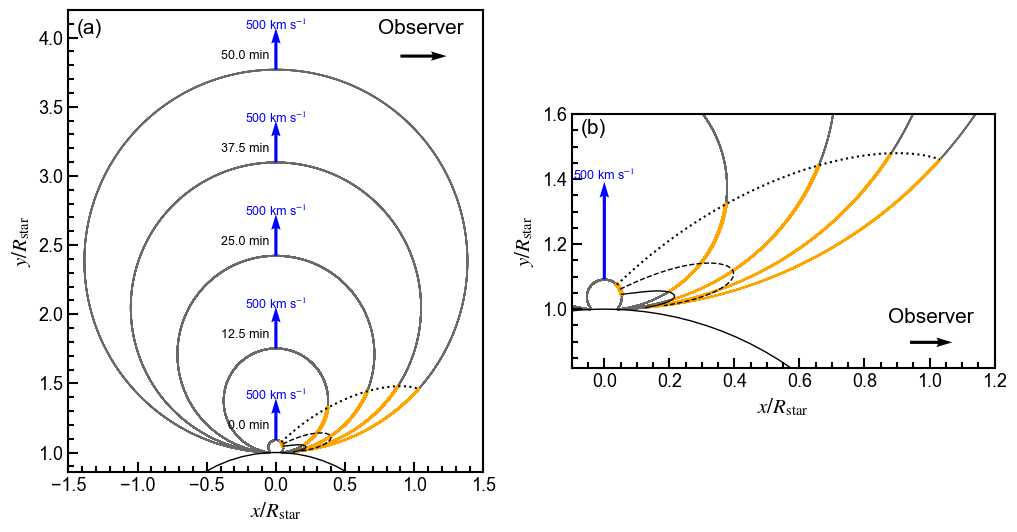}
\caption{
Results from a one dimensional hydrodynamic simulation of the flow along the expanding magnetic loop (pseudo 2D MHD model; Section \ref{sec:3-3}) described in \citet{2024ApJ...963...50I} and Ikuta et al. in prep.. 
(a) Schematic representation of the magnetic loop expanding at a constant velocity (depicted as a loop-like structure at different times). 
The observer's direction is indicated by the black axis. 
The loop's footpoints are connected to the star, located at the stellar limb. The direction of the eruption is perpendicular to the line of sight (LOS).
The prominence material in the loop is indicated with the orange color. 
The Lagrangian trajectory of plasma at the bottom, middle, and top of the original prominence are indicated with black solid, dashed, and dotted lines.
(b) Expanded picture of panel (a).
\Add{The initial condition of the magnetic loop and explanations of parameters are described in Appendix \ref{app:setup}.}
}
\label{fig:ikuta-model}
\end{figure*}

In this model, the shape and time variation of the magnetic field configuration is assumed on the basis of an approximate version of the self-similar magnetohydrodynamic outflow for a typical model of CME (e.g., \citealt{1997SoPh..175..601D,2017ApJ...847...27A}; see also theoretical self-similar CME model developed by \citealt{1984ApJ...281..392L,1998ApJ...493..460G}). 
Figure \ref{fig:ikuta-model} shows the assumed magnetic loop expansion model.
The plasma velocity along an expanding magnetic loop (due to gravity and gas pressure) is simulated, whereas the plasma velocity normal to magnetic field line is assumed to be given based on a model of expanding loop whose initial velocity is a free parameter which is determined from comparison with observations. 
Therefore, we can call this a ``pseudo 2D MHD model" for prominence/filament eruption. 
This model nicely \Add{captures} the motion of the erupting prominence/filament which is decelerated by the gravity after ejection, and \Add{falls} along the erupting magnetic field line in the case of \Add{a} solar erupting prominence/filament \citep{2024ApJ...963...50I}. %

Here we assume the configuration of a bipolar magnetic field by two line currents and introduce bipolar coordinates defined by this bipolar magnetic configuration as in \cite{1980SoPh...66...61S}.
The parameters of the loop configuration are
(1) the half distance between the two line poles, which corresponds to the size of an active region at the loop bottom, $a$,
(2) the initial height of the loop top, $y_{\rm int}$, and 
(3) the initial normal velocity of the loop top, $V_{\rm top}$.
We also assume that there is a cool and dense prominence with uniform temperature of $10^4$ K and density of $10^{-13}$ g cm$^{-3}$ in a localized region in the loop where the curvilinear coordinate parameter $v$ is between $v_{\rm min}$ and $v_{\rm max}$ (0 $<$ $v_{\rm min}$ $<$ $v$ $<$ $v_{\rm max}$ $<$ 3) at the initial magnetic loop. 
The assumed density is consistent with solar observations ($10^{-14}$--$10^{-12}$ g cm$^{-3}$; \citealt{1986NASCP2442..149H}).
Then, the uniform pressure is given by 0.166 dyn cm$^{-2}$. 
For the outside of the prominence in the magnetic loop, we assume that there is a hot corona with a temperature of $10^6$ K and density of 10$^{-15}$ g cm$^{-3}$, whose pressure is balanced with the prominence \citep[see][for discussion of these assumptions]{2024ApJ...963...50I}. 

As a result of this numerical simulation, we can obtain the time evolution of the temperature, density, and velocity of the plasma inside the 1D loop.
To model the effects of different observer perspectives, we introduce a viewing angle parameter $\theta$, which represents the angle between the observer’s LOS and the direction of the ejection. 
For simplicity, we only consider the model viewed from the plane of the 2D magnetic loop. 
Based on the results obtained in Section \ref{sec:3-2}, we adopted \(\theta = 90^\circ\) (see Figure \ref{fig:ikuta-model}). 
While it is possible to slightly change the angle, we did not adopt this approach because, due to the nature of the model, even a small tilt in the angle causes the falling plasma to be projected onto the disk surface, resulting in an absorption profile. 
However, in future work, we will conduct a broader parametric study.

To compare the model with observations, we calculated the LOS velocity component at each point of the loop. 
Then, we summed up the mass of the prominence material (plasma with temperatures less than \(3 \times 10^4 \, \text{K}\)) and derived the mass distribution as a function of velocity, simulating the pseudo H$\alpha$ spectrum under the optically thin condition. Further radiative transfer calculations were not conducted in this study. 
Instead, we scaled this modeled dynamic spectrum and compared only its velocity variations with the observations of EK Dra.







\begin{figure*}
\epsscale{1.0}
\plotone{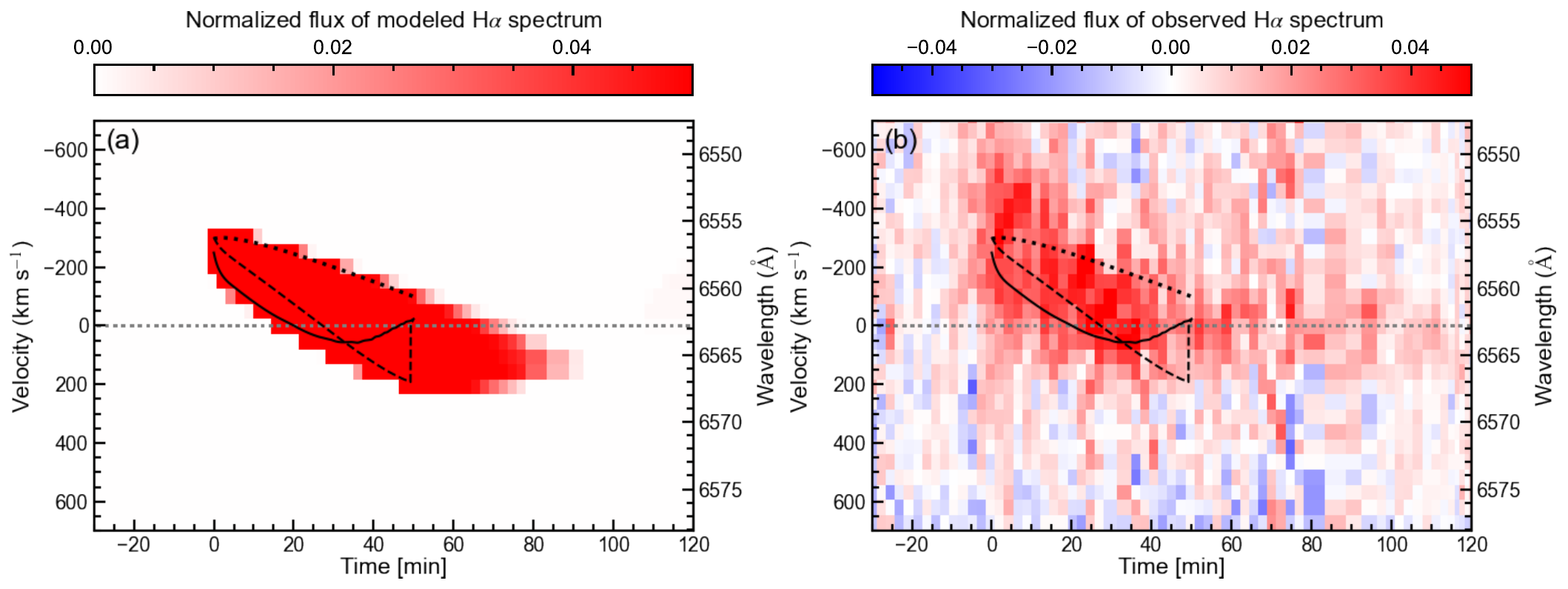}
\caption{
\Add{Comparison between modeled and observed dynamic spectrum of H$\alpha$ line from a one dimensional hydrodynamic simulation of the flow along the expanding magnetic loop (pseudo 2D MHD model; Section \ref{sec:3-3}).}
\Add{(a)} Simulated dynamic spectrum of the H$\alpha$ line from the prominence material.
Each black line corresponds to the velocity evolution of each Lagrangian trajectory of plasma in panel (a) \Add{of Figure \ref{fig:ikuta-model}}.
\Add{(b)} Observed dynamic spectrum of the H$\alpha$ line during the giant prominence eruptions on EK Dra, as reported in \textsf{Paper \hyperref[2022ApJ...926L...5N]{I}}.
}
\label{fig:ikuta-model-ds}
\end{figure*}

As a result of the parameter survey, the optimal parameters that best match the observations were determined as follows: ($a$, $y_{\rm int}$, $V_{\rm top}$, $v_{\rm min}$, $v_{\rm max}$) = ($2.5 \times 10^4$ km, $6.0 \times 10^4$ km, $5.0 \times 10^2$ km s$^{-1}$, 0.35, 0.75).
\Add{The assumed velocity of the magnetic loop is plausible, as the typical Alfvén speed in stellar coronae is on the order of 1000 km s$^{-1}$ \citep{2011LRSP....8....6S}.}
The orange line in Figure \ref{fig:ikuta-model} (\Add{top}) illustrates the assumed location of prominence material (based on the above parameters $v_{\rm min}$ and $v_{\rm max}$) in the expanding magnetic loop, and its time evolution. 
As observed in \cite{2024ApJ...963...50I} and in Figure \ref{fig:ikuta-model}, we assume that the prominences are predominantly distributed on one wing of the magnetic loop (that is expanding towards Earth), which is often observed in the Sun.
In this setup, the prominence is initially ejected, seemingly pushed by the expanding loop, but is then observed to fall back toward the star along the loop. This behavior is consistent with the filament eruption case described by \cite{2024ApJ...963...50I}.
Figure \ref{fig:ikuta-model-ds} shows the comparison between the modeled pesudo-H$\alpha$ dynamics spectrum and the observed H$\alpha$ dynamics spectrum.
The overall features of the observation and model looks similar with each other, indicating that our model successfully reproduces the observed behaviors to some extent.
It should be noted, however, that the final velocities in our model show a redshifted component offset by several tens of km s$^{-1}$, while is not seen in the observations.
In addition, the very fast and faint component of $>$ 600 km s$^{-1}$ was not well reproduced in this model and the model in Section \ref{sec:3-2}.
This discrepancy may suggest minor deviations from the model's ability or just a change in the H$\alpha$ luminosity of the prominence, which will be further discussed in Section \ref{sec:4}. 
Further adjustments to the simulation parameters may be required to refine the alignment with observed data and to explore the underlying causes of these velocity offsets, comparing with a solar prominence eruption (Ikuta et al. in prep.).

Figure \ref{fig:mcmc-velocity} compares the velocity and height of prominences in the 1D HD model with those in the 1D free-fall model discussed in Section \ref{sec:3-2}. According to this comparison, both models agree within the range of diversity in height and velocity changes predicted by the 1D HD model. The prominence center velocities in the 1D HD model are slightly slower, which consequently results in slightly lower achieved heights. This outcome supports the validity of both models.

\section{Discussion}\label{sec:4}

\subsection{Starspots and Magnetic Field Topology from Light Curve Inversion and (Zeeman) Doppler Imaging} \label{sec:4-1}




Here, we compare the spot maps from TESS light curve modeling with those derived using Doppler Imaging (DI). 
The TESS light curve modeling results depicted in Figure \ref{fig:starspots}(b) reveals that during a prominence eruption, all three hypothesized spots were located on the hemisphere visible from Earth. 
These are designated as spots A, B, and C and their properties are summarized in Table \ref{tab:spotparams}.
The result indicates their presence primarily in mid latitudes (25, 25 and 48 degrees), although we should keep this in mind that the light curve modeling has inherent limitations such as the degeneracy between latitude and spot area.
Additionally, these spots were positioned near the limb as viewed from Earth ($\sim$20-30 deg from the limb). 
During the observation, spots A and B maintained a significant area, whereas spot C was notably smaller. 
Note that these are temporally evolving spots, and the minimal area of spot C at this time does not imply it is always small.

The DI results shown in Figure \ref{fig:starspots}(a) reveal large, dark regions suggestive of significant spots at mid-latitudes (the latitude of 30-40 degree).
This is consistent with those observed in the TESS light curve modeling in Figure \ref{fig:starspots}(b), although the DI map shows more extended spot distributions.
This consistency between the two methods validates their spot configuration results. 
While the TESS light curve inversion traces time variations, Doppler Imaging (DI) does not, resulting in an average spot distribution over a month. 
Therefore, the two methods are not necessarily identical or completely equivalent, but the obtained consistency may suggest that the spot location (at least the active latitude) does not change significantly within a month. 
\Add{This indication is consistent with the fact that the lifetime of giant starspots on solar-type stars extend to several months to a year \citep{2019ApJ...871..187N,2020ApJ...891..103N}.}

In contrast to the TESS light curve inversion, the DI results also suggest the presence of a substantial polar spot (see Section \ref{section:2-1}).
This is a feature undetectable by TESS light curve modeling as stellar rotation does not affect the brightness variation due to such polar spots. 
Thus, the discrepancies between the two methods in terms of polar spots are not contradictions but rather demonstrate how our complementary observations with DI can compensate for the limitations of TESS light curve modeling. 
A detailed global mapping comparison will be addressed in our future work (Ikuta et al. in prep.). In this paper, we focus on the spot locations during a prominence eruption as a snapshot in time.

Figure \ref{fig:magmap} shows the magnetic field distribution obtained via Zeeman Doppler Imaging (ZDI).
The ZDI's spatial resolution is at least 20-25 deg (see, Appendix \ref{app:a}).
Also, we need to keep in mind that ZDI \Add{maps tend} to cancel out small-scale magnetic fields.
So, it is important to recognize that the map primarily reflects the distribution of large-scale magnetic fields, rather than local spot's magnetic fields.
The radial magnetic field at the poles displays an almost negative unipolar configuration, suggesting that the large polar spots in the DI map (Figure \ref{fig:starspots}(a)) are either associated with a unipolar magnetic field or situated within a large-scale dipole or open magnetic field structure, like a coronal hole.
To verify this, we performed a simple field extrapolation using a potential field source surface model (Figure \ref{fig:potential-field}) and found that a substantial part of the unipolar negative field at the polar region hosts open magnetic field lines.
On the other hand, near mid-latitudes, the substantial spots labeled A and B are located near a polarity inversion line (PIL).
In Figure \ref{fig:starspots}(c), in more detail, the large starspot B region on the western limb is clearly located on the PIL. Although starspot A region on the eastern limb is not precisely on the PIL, considering the spatial resolution, it might be associated with the PIL.
The existence of these spots in different magnetic environments implicates their potential role in influencing the likelihood of CME launches or the presence of pre-flare giant prominences, which is discussed in Section \ref{sec:4-2}.


\subsection{The Origin of Gigantic Prominence Eruption from EK Dra on April 10th, 2022}\label{sec:4-2}

Here we explore the relationship in the locations of starspots, magnetic fields, and prominence eruptions.
We focus on the possibility that the superflares and major prominence eruptions are linked to large starspots observed in EK Dra, while prominence eruptions could occur in quiet regions, like solar cases.
We begin by synthesizing information regarding the distribution of prominence eruptions, large starspots, and the magnetic field configuration:
\begin{itemize}
    \item Eruptive prominences are \Add{likely ejected} from the footpoints located approximately 12-16 degrees from the stellar limb (Section \ref{sec:3-2} and \ref{sec:3-3}).
    \item These footpoints coincide with a polar region starspot hosting unipolar large-scale magnetic fields and open coronal fields, as well as mid-latitude large spots (A \& B) hosting large-scale polarity inversion lines (PIL) and closed coronal feilds (Section \ref{sec:4-1}).
\end{itemize}
The consistency suggests that the massive prominence eruption on April 10th, 2022 can be associated with the observed giant starspots.

In the case of the Sun, large prominences are often observed along polarity inversion lines. 
Based on this, it can be expected that mid-latitude spots with extensive polarity inversion lines could be the sites of the eruptive prominence. 
It should be also noted that these regions would have an overlying closed magnetic field as seen in the coronal field extrapolation presented in Figure \ref{fig:potential-field}. 
An overlying large-scale dipolar magnetic field can suppress the ejected structure, as indicated in the numerical studies of \citet{2018ApJ...862...93A}. These authors demonstrated that, for a solar-type star with a 75-G dipolar field, escaping CMEs would have kinetic energies $\gtrsim 3 \times 10^{32}$ erg. 
Because the estimated kinetic energy of the prominence eruption $5.8_{-4.0}^{+12.8}\times 10^{34}$ erg derived in \textsf{Paper \hyperref[2022ApJ...926L...5N]{I}} is much higher than this threshold (see Table \ref{tab:energy-scale}), the massive prominence eruptions and related CMEs might not be suppressed significantly (see, e.g., Strickert et al. 2024, submitted).
Note that the observed average magnetic field of EK Dra of 120 G is slightly stronger than the 75 G and its magnetic topology is not a dipole, so the applications of study by \citet{2018ApJ...862...93A} could be not so straightforward and there could be some differences for this threshold. 
A future modeling effort based on the observed magnetic field of EK Dra is required for the exact evaluation.

The polar spot we detected is located in a region of almost unipolar large-scale field, which at first sight could indicate that 
it is not likely related to the pre-existing \Add{gigantic} prominence. However, because the small-scale fields are not recovered with the ZDI technique, it is possible that polarity inversions caused by, e.g., bipolar regions, could take place within the polar spot. 
Therefore, if the giant prominence originated within the polar spot region, where the stellar coronal field is open, 
the prominence could erupt without suppression by the closed dipolar magnetic fields (see, e.g., Strickert et al 2024, submitted).
Furthermore, although subtle, it appears that the edge of the polar spot is in contact with the PIL, which might be related to the pre-existing polar prominence.
Future studies should include numerical simulations based on these magnetic field maps to validate each scenario.

In the above context, the assumption of a pre-existing prominence or filament along \Add{the} PIL suggests that, following its eruption, post-flare dimming or enhancement should be anticipated at the center of the H$\alpha$ line. The absence of clear evidence for this phenomenon may imply either the absence or minimal size of the pre-existing prominence or filament. 
The effect may be relatively minor \citep{2022ApJ...939...98O} and it could be simply challenging to detect unambiguously.
However, if a pre-existing prominence or filament be located at low-mid latitudes, the associated effects will be detected shifted by the rotational velocity. 
Looking at this Doppler effect is challenging at a spectral resolution of R$\sim$2000, \Add{however} future high-dispersion spectroscopy may potentially allow for the localization of such pre-flare features.

\subsection{Time Evolution of the Erupted Prominence: Can a CME Occur?}\label{sec:4-3}

Here, we examine the temporal evolution of prominence eruptions based on the simplified models presented in Sections \ref{sec:3-2} and \ref{sec:3-3}, and explore whether and how the eruption ultimately led into a stellar CME.

First, as shown in Figure \ref{fig:mcmc-velocity} and described in Section \ref{sec:3-2}, the one-dimensional free-fall model provides insights into the temporal variations in velocity and height of the prominence, with results derived from two distinct fittings to the H$\alpha$ line data. 
These analyses suggest that the prominence's center of mass was lifted to between 0.5 and 1 stellar radii over a period of 40 minutes to one hour before it eventually fall back to the star. 
Moreover, in Section \ref{sec:3-3}, a one dimensional hydrodynamic simulation of the flow along the expanding magnetic loop (pseudo 2D MHD model), without any radiative transfer model, also yields solutions that slightly indicate redshift at the end.
A key issue observed is that this particular prominence eruption ultimately disappeared near zero velocity without exhibiting any redshift in the H$\alpha$ line. 
We found that the anticipated timing for the onset of detectable redshift, about one hour after the event, coincides with the H$\alpha$ line flux approaching near zero. 
This suggests a plausible scenario where the prominence, having possibly become too tenuous or heated (e.g., \citealt{2021EP&S...73...58S} explain the solar cases), was no longer visible in the H$\alpha$ line at the onset of detectable redshift.

Both models presented in Sections \ref{sec:3-2} and \ref{sec:3-3} support the scenario in which the prominences erupt near the limb and project at a certain angle in the vertical direction, predominantly in the direction of the Earth. 
Such tilted trajectories of prominence eruptions are commonly observed on the Sun \citep[e.g.,][]{2019SunGe..14...95S}, and thus, there is no contradiction in these findings.
This tilt can be related the surrounding magnetic environment and/or an uneven distribution of prominences within magnetic loops (cf. Figure \ref{fig:ikuta-model}).
There is a possibility that strong magnetic fields or large-scale magnetic fields (such as strong toroidal magnetic fields) of EK Dra may be involved (see Figure \ref{fig:magmap}), so verification by future 3D ejection models will be necessary.
We emphasize that the obtained evolution likely represents one of the most well-characterized events in the study of stellar prominence eruptions across G/K/M-type stars.

As discussed in \textsf{Paper \hyperref[2022ApJ...926L...5N]{I}}, it has been suggested that this prominence eruption could have evolved into a CME for various reasons: 
(i) The blueshift velocity slightly exceeded the escape velocity.
(ii) Extrapolating from the relationship between the velocities of solar prominences and CMEs, the outer-layer coronal velocity is expected to exceed the escape velocity by enough.
(iii) The criteria for the occurrence of a CME, based on the relationship between the length scale and velocity of solar prominence eruptions, were significantly surpassed.
In this \textsf{Paper II}, the results from Sections \ref{sec:3-2} and \ref{sec:3-3} imply that the center of mass of the prominence might have fallen back towards the stellar surface. 
However, this only reflects the motion of the center of mass. 
Typically, prominences, like those on the Sun, are spatially extensive, and considering sufficient velocity dispersion, parts of these structures are expected to exceed the escape velocity, potentially leading to a CME. 
Indeed, the extended area for our eruptive prominence \Add{was} estimated in \textsf{Paper \hyperref[2022ApJ...926L...5N]{I}} (cf. large area was also inferred for prominences on a young solar-type star, V530 Persei; \citealt{2020A&A...643A..39C}).
Furthermore, one-dimensional HD calculations based on solar physics in Section \ref{sec:3-3} support the scenario where, even if the prominence is falling, the magnetic loops continue to expand beyond the escape velocity. This substantiates the assertions made in \textsf{Paper \hyperref[2022ApJ...926L...5N]{I}}, significantly reinforcing this model.

Finally, we discuss the possible role of centrifugal force in the evolution of prominences. 
In rapidly rotating solar-type stars with rotation periods of less than one day, prominences supported by centrifugal force due to the stellar rotation are often observed (referred to as ``slingshot" prominences; \citealt{1989MNRAS.236...57C,2000MNRAS.316..699D,2006MNRAS.365..530D,2020A&A...643A..39C}).
Since EK Dra is a rapidly rotating star, though slower than stars having slingshot prominences, centrifugal force may aid in the ejection of prominences. The ratio of centrifugal force to gravity is given by $\omega^2 (h + R_{\rm star})^3 / G M_{\rm star}$, where $h$ is the height above the stellar surface. This ratio is 0.01 at $h = 1R_{\rm star}$, 0.04 at $h = 2R_{\rm star}$, 0.33 at $h = 5R_{\rm star}$, and 1.1 at $h = 8R_{\rm star}$. Both models in Section \ref{sec:3} suggest that the prominence height from the stellar surface reaches about 0.5 to 1 stellar radius. This indicates that, within the framework and results of both eruption models, the contribution of centrifugal force is negligible, being only a few percent of the gravity. However, as mentioned earlier, some erupted prominences are expected to reach higher heights, and could be significantly accelerated by centrifugal force, especially from around 8 stellar radii.

\begin{deluxetable}{cccccccc}
\label{tab:energy-scale}
\tablecaption{Summary of energy distribution of  superflares on EK Dra estimated in \textsf{Paper \hyperref[2022ApJ...926L...5N]{I}}.}
\tablewidth{0pt}
\tablehead{
\colhead{Event ID} & \colhead{$fE_{\rm mag}$} & \colhead{$E_{\rm WLF}$}  & \colhead{$E_{\rm X}$} & \colhead{$E_{\rm kin}$}  \\
\colhead{} & \colhead{[10$^{33}$ erg]} & \colhead{[10$^{33}$ erg]} & \colhead{[10$^{33}$ erg]} & \colhead{[10$^{33}$ erg]} 
}
\startdata
E1 & 34 & 1.5$_{\pm 0.1}$  & -- & $58^{+128}_{-40}$  \\
(norm.)$^{\dagger}$ & (1) & (0.043) & (--) & (1.7$^{+3.7}_{-1.2}$) \\
E2 & 54 & 12.2$_{\pm 0.2}$  & $12.3_{-0.5}^{+0.3}$-$16.7_{-0.7}^{+0.4}$ & $1.2^{+2.7}_{-0.8}$  \\
(norm.)$^{\dagger}$ & (1) & (0.228) & (0.230-0.312) & (0.022$^{+0.050}_{-0.015}$) \\
E4 & 29 & 2.0$_{\pm 0.1}$  & -- & 0.35$^{+1.40}_{-0.30}$  \\
(norm.)$^{\dagger}$ & (1) & (0.068) & (--) & (0.012$^{+0.048}_{-0.010}$) 
\enddata
\tablecomments{
The table data is taken from \textsf{Paper \hyperref[2022ApJ...926L...5N]{I}}. 
$E_{\rm WLF}$ is white-light flare energy, $E_{\rm X}$ is X-ray flare energy, and $E_{\rm kin}$ is kinetic energy of eruptive prominence.
The free magnetic energy $fE_{\rm{mag}}$ is estimated by the equation of $fE_{\rm{mag}} = \frac{1}{8\pi} f B^2 L_{\rm{spot}}^{3}$ \citep[e.g.,][]{2013PASJ...65...49S}, where a filling factor $f$ is 0.1 and averaged surface magnetic field strength of spot is 1000 G.
Not that in \textsf{Paper \hyperref[2022ApJ...926L...5N]{I}} the free magnetic energy was mistakenly estimated under the $f$ value of unity, which significantly overestimated the free magnetic energy by an order of magnitude. Here we correct the values.
$^{\dagger}$The columns ``norm." mean the values normalized by the free magnetic energy $fE_{\rm{mag}}$.
}
\end{deluxetable}

\begin{figure*}
\gridline{
\fig{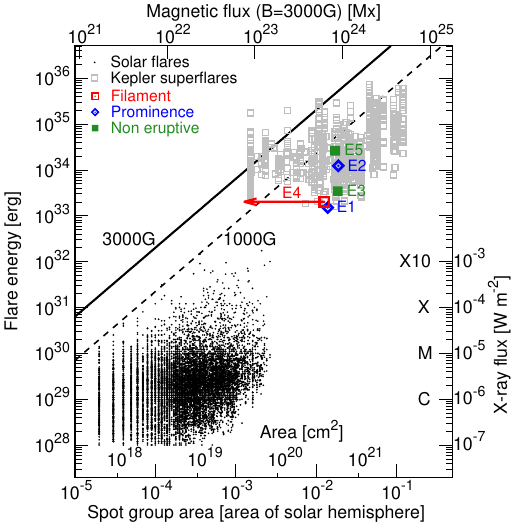}{0.5\textwidth}{\vspace{0mm} (a)}
\fig{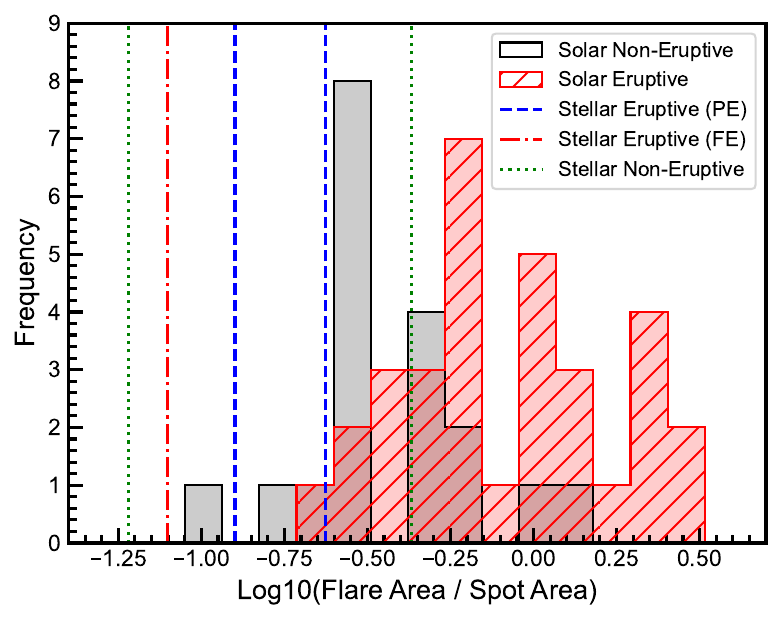}{0.5\textwidth}{\vspace{0mm} (b)}
}
\caption{
(a) The flare energy (\(E_{\rm{flare}}\)) as a function of spot group area (\(A_{\rm{spot}}\)) of solar flares and superflares on solar-type stars.
The lower axis is the area of spot group in the unit of the solar hemisphere ($1/2\times A_{\sun}\sim 3\times 10^{22}$cm$^{2}$).
The top horizontal axis is the magnetic flux of spots under the assumption of $B=3000$ G.
The left vertical axis shows the bolometric energy released by each flare.
The right vertical axis GOES X-ray class where we assume that bolometric energies of $10^{29}$, $10^{30}$, $10^{31}$, and $10^{32}$ erg scales to the GOES X-ray class of C, M, X, and X10, respectively.
The black solid and dashed lines is the relationship between $E_{\rm flare}$ and $A_{\rm spot}$ for $B=3000$ G and 1000 G, respectively.
The relation assume that the maximum flare energy against a given spot area can be explained by the total magnetic energy, i.e., $E_{\rm{flare}} = \frac{1}{8\pi} f B^2 A_{\rm{spot}}^{3/2}$, where f is a filling factor \citep[e.g.,][]{2013PASJ...65...49S}.
Here we did not add the error bars in data, because the xy axis is in log scale and the error range is too small to be plotted.
(b) Comparison of the histograms for eruptive and non-eruptive events as a function of the ratio of flare area to spot area for the Sun and EK Dra.
Solar data are taken from \cite{2017ApJ...834...56T} who derived the flare ribbon size from temporally stacked urtraviolet 1600 {\AA} images.
``FE" and ``PE" means filament and prominence eruptions, respectively.
}
\label{fig:spot-flare}
\end{figure*}

\subsection{Relationship between Eruptive and Non-Eruptive Superflares in Comparison with Starspot Area}\label{sec:4-4}

Finally, we explore the relationship between spot size and flare magnitude by comparing solar flares to those observed on EK Dra. 
It is known that, for solar flares and CMEs, when the flare's \Add{length} scale significantly exceeds the associated sunspot's \Add{length} size, the flare tends to be eruptive rather than being suppressed by the magnetic field. 
Statistical analysis supports that larger flare length scales relative to sunspot length scales are more prone to result in CMEs \citep{2017ApJ...834...56T,2023ApJ...958..104K}. 
Given the scale-free nature of MHD, these may also apply to large stellar superflares.

Here, we propose the introduction of parameters transferred from solar research into the study of stellar eruptions: the flare energy  relative to spot magnetic energy (Figure \ref{fig:spot-flare}(a)), or the flaring area relative to spot area (Figure \ref{fig:spot-flare}(b)).
Figure \ref{fig:spot-flare}(a) compares the flare energy-spot size relation between solar flares, flares on Kepler solar-type stars \citep{2020arXiv201102117O}, and recent observations from EK Dra (\textsf{Paper \hyperref[2022ApJ...926L...5N]{I}}). 
\cite{2013PASJ...65...49S} proposed that the magnetic energy available within the spot determines the maximum flare magnitude, as in the lines in Figure \ref{fig:spot-flare}, and flares close to this line can release a large portion of the stored free energy.
We plotted events from EK Dra as either non-eruptive\footnote{Here, ``non-eruptive" is defined as events where no blueshift was observed. However, considering LOS uncertainties, we cannot rule out that these were truly non-eruptive events \Add{only from this definition. If prominences erupt in a direction nearly perpendicular to the LOS, they will not be detected as blueshifts even if eruption occur.} } or eruptive to investigate potential biases suggested in this graph. 
However, the limited sample of five cases shows no clear distinction. 
Given uncertainties such as which spot hosted each event\Add{,} or LOS ambiguities that might obscure observed eruptions, a larger dataset is necessary to discern trends. 
Nonetheless, the proximity of all EK Dra's events to the line in the graph suggests that a large portion of the magnetic energy in the spots was likely released for these superflares.
Table \ref{tab:energy-scale} presents a comparison of the free magnetic energy\footnote{In \textsf{Paper \hyperref[2022ApJ...926L...5N]{I}}, the free magnetic energy was mistakenly estimated under the $f$ value of unity, although the text says that we assumed $f=0.1$, which led to the incorrect conclusion that ``Both the kinetic and radiation energies are much smaller than the stored magnetic energy of the active regions/starspots". Here we have recalculated the correct values and revised the discussion.}, flare radiation energy, and kinetic energy of prominence eruptions. 
Specifically, assuming a filling factor $f$ of 0.1, the event E2 converts 23$_{\pm 0.3}$\% of the flare's magnetic energy into radiation energy, whereas the event E1 converts nearly all (170$^{+370}_{-120}$\%) of the flare's magnetic energy into kinetic energy. These findings indicate that Events E1 and E2 represent extreme cases among the observed starspots.

Further, Figure \ref{fig:spot-flare}(b) presents a histogram comparing the ratio of flare length scale to sunspot length scale for solar flares \citep[data taken from][]{2017ApJ...834...56T}, with the eruptions from EK Dra overplotted.
\cite{2017ApJ...834...56T} reported that this ratio statistically differs between eruptive and non-eruptive events for solar flares. 
{Among EK Dra's events,} we found that even smaller-scale flares relative to spot size were eruptive.
Although the direct comparison with solar statistics by \cite{2017ApJ...834...56T} is not straightforward because they do not use the same flare size measurements, 
\Add{we plotted this as a reference, and we can at least say that overall both values are consistent.}
Although the current number of the sample is only five, this \Add{result} may mean that the threshold may not necessarily hold for active stars, or that eruptive events on active stars can emerge not from the dominant spots but possibly from clusters of smaller spots.
The latter possibility is partly suggested by the fact that the each spot size with TESS light curve modeling ((0.22-1.02)$\times10^{10}$ cm, as in Section \ref{section:2-2}) is smaller than that obtained by global TESS light curve \citep[2.1$\times10^{10}$ cm,][]{2024ApJ...961...23N}.

\section{Summary and Conclusion}

In \textsf{Paper \hyperref[2022ApJ...926L...5N]{I}}, we reported the discovery of a gigantic prominence eruption on the young solar-type star, EK Dra. This prominence, occurring outside the stellar disk, is constrained by well-defined velocity variations over time, providing an excellent dataset for investigating its dynamics. 
In the present paper, \textsf{Paper II}, our first objective is to estimate the dynamical evolution of this stellar prominence eruption. We employed a simple one-dimensional free-fall model (Section \ref{sec:3-2}) and 
a one dimensional hydrodynamic simulation of the flow along the expanding magnetic loop (pseudo 2D MHD model; Section \ref{sec:3-3}) to infer the direction of ejection, changes in height, and the possible development of a magnetic loop.
Furthermore, our second objective is to explore the origins of the prominence eruption and its relationship with the small-scale and large-scale magnetic field. 
For this purpose, we analyzed the TESS light curve  (Section \ref{section:2-2}) and spectropolarimetric observation data obtained concurrently with the prominence observations (Section \ref{section:2-1}). 
These analyses allowed us to estimate the maps of starspots and the large-scale magnetic field, and to examine their possible relationship with the prominence eruption. 
We suggest the following scenario for this event:
\begin{enumerate}
    \item Location: The massive stellar prominence eruption is estimated to have originated approximately 12-16 degrees from the limb and was ejected at an angle of about 15-24 degrees relative to the line of sight (Section \ref{sec:3-2}). This spatial evolution can be explained even by an expanding magnetic loop structure (Section \ref{sec:3-3}, Figure \ref{fig:ikuta-model}).

    \item Evolution: The center of mass of the prominence is expected to have reached its peak height about 40 minutes to an hour after eruption, followed by an anticipated descent (Section \ref{sec:3-2}, Figure \ref{fig:mcmc-velocity}). The timing coincides with the disappearance of the H$\alpha$ line intensity. This can explain the puzzle in \textsf{Paper \hyperref[2022ApJ...926L...5N]{I}} (see Section \ref{sec:1}) by \Add{suggesting} that the prominence may have disappeared at zero velocity merely because it became invisible in \Add{the} H$\alpha$ line.  
    Our one dimensional hydrodynamic simulation of the flow along the expanding magnetic loop (pseudo 2D MHD model) indicate that even if the center of mass begins to fall, the loop itself continues to expand and can evolve into a CME (Section \ref{sec:4-3}, Figure \ref{fig:ikuta-model}).

    \item Origin: At the time of the prominence eruption, there were several large starspots at mid latitudes on the limb, and a large starspot near the \Add{pole} (Section \ref{sec:4-1}, Figures \ref{fig:starspots}, \ref{fig:magmap}), consistent with the observed location of the base of the prominence eruption (Section \ref{sec:4-2}). The large starspot at mid-latitude, located on the polarity inversion line, is likely to be the source of the massive prominence, while we cannot rule out the possibility that the polar spot with single polarity might be the source of the successful eruption.

\end{enumerate}
There have been no studies that have extensively compared the direction of ejection and magnetic field environments based on observational data. 
Therefore, this is one of the most detailed studies in the field of stellar CME research, estimating the evolution and environmental context.
These results inferred from the representative young Sun-like star provide valuable insights into the dynamic processes that likely influenced the environments of the early Earth, Mars, Venus, and young exoplanets.

To further investigate the correlation with spots, we conducted an analysis comparing the ratio of spot size to flare size, a criterion used on the Sun to differentiate between eruptive and non-eruptive events, across all solar flares and the superflares observed on EK Dra. While the observed stellar flares likely released a significant amount of magnetic energy, any trends were not consistently evident. 
Given the current small sample size, it is expected that increasing the number of events in future studies may reveal some trends.

Finally, we summarize the future direction of this series of papers. Future work will focus intensively on comparing spots and active regions observed across multiple wavelengths. 
While the current study analyzed the configuration of spots at a snapshot, we will investigate the temporal changes in their global distribution and variations in active regions as observed in X-rays and H$\alpha$. 
Additionally, we plan to conduct more detailed dynamical modeling of stellar prominences in comparison with solar prominence observations and models, and studies estimating the parameters of prominences through radiative transfer modeling. 
These efforts will allow us to analyze this event in greater detail.

\section*{Acknowledgment}

\Add{We would like to thank the referee for their valuable comments and suggestions, which have helped to improve the quality of our manuscript.}
The authors thank Dr. Juli\'{a}n D. Alvarado-G\'{o}mez for fruitful comments on DI and ZDI maps. 
This work was supported by JSPS (Japan Society for the Promotion of Science) KAKENHI Grant Numbers 21J00316 (K.N.), 24K17082 (K.I.), 20K04032, 20H05643, 24K00685 (H.M.), 21H01131 (H.M., K.I., and K.S.), 24H00248 (K.N., K.I., and H.M.), and 24K00680 (K.N., H.M., and K.S.). 
V.S.A. was supported by the GSFC Sellers Exoplanet Environments Collaboration (SEEC), which is funded by the NASA Planetary Science Division’s Internal Scientist Funding Model (ISFM), NASA NNH21ZDA001N-XRP F.3 Exoplanets Research Program grants and NICER Cycle 2 project funds and NICER DDT program.
A.A.V. acknowledges funding from the European Research Council (ERC) under the European Union's Horizon 2020 research and innovation programme (grant agreement No 817540, ASTROFLOW).
Y.N. was supported from the NASA ADAP award program Number 80NSSC21K0632.
\Add{P.H. acknowledges support from the grant 22-34841S of the Czech Science Foundation and he was also supported by the program ``Excellence Initiative--Research University" for the years 2020--2026 at the University of Wrocław, project No. BPIDUB.4610.96.2021.KG. P.H. and J.W. were supported by the funding RVO:67985815}
A.A.A. acknowledges Bulgarian NSF grant No.KP-06-N58/3 (2021).
S.V.J. acknowledges the support of the DFG priority program SPP 1992 ``Exploring the Diversity of Extrasolar Planets" (JE 701/5-1).
The spectroscopic data used in this paper were obtained through the program 22A-N-CN06 (PI: K.N.) with the 3.8m Seimei telescope, which is located at Okayama Observatory of Kyoto University.
This paper includes data collected with the TESS mission, obtained from the MAST data archive at the Space Telescope Science Institute (STScI). Funding for the TESS mission is provided by the NASA Explorer Program. STScI is operated by the Association of Universities for Research in Astronomy, Inc., under NASA contract NAS 5-26555. 
Some of the data presented in this paper were obtained from the Mikulski Archive for Space Telescopes (MAST) at the Space Telescope Science Institute. The specific observations analyzed can be accessed via \dataset[10.17909/xgds-j146]{https://doi.org/10.17909/xgds-j146}.
The authors acknowledge ideas from the participants in the workshop ``Blazing Paths to Observing Stellar and Exoplanet Particle Environments" organized by the W.M. Keck Institute for Space Studies.

\facilities{Seimei telescope, T\'{e}lescope Bernard Lyot (TBL), Transiting Exoplanet Survey Satellite (TESS)}

\software{\textsf{astropy} \citep{2018AJ....156..123A} , \textsf{IRAF} \citep{Tody1986}, \textsf{PyRAF} \citep{2012ascl.soft07011S}}


\appendix

\section{\Add{Initial configurations and parameters of the pseudo 2D MHD model}}\label{app:setup}

\begin{figure}
\epsscale{0.5}
\plotone{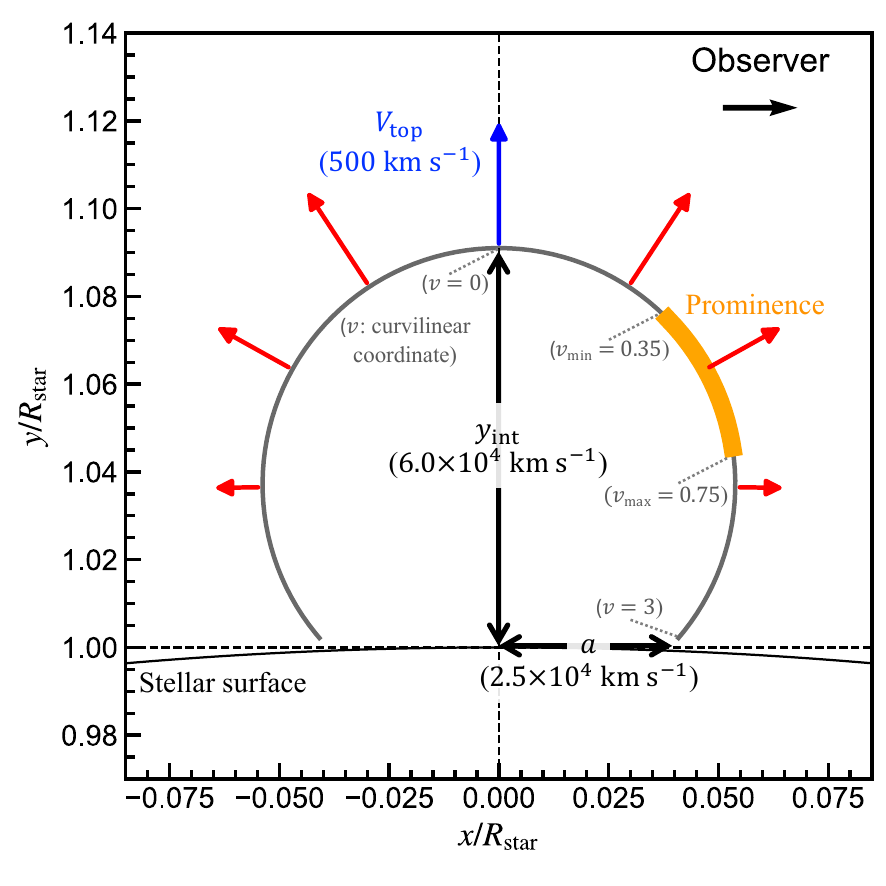}
\caption{
\Add{Initial configurations and parameters of a one dimensional hydrodynamic simulation of the flow along the expanding magnetic loop (pseudo 2D MHD model; Section \ref{sec:3-3}) described in \citet{2024ApJ...963...50I} and Ikuta et al. in prep..
The initial position of the prominence is indicated by the thick orange line. 
In our model, specifying the velocity parameter at the loop apex, \( V_{\text{top}} \) (blue arrow), automatically determines the velocity distribution in the other parts of the magnetic loop (red arrows).}
}
\label{fig:ikuta-model-setup}
\end{figure}

\Add{
Figure \ref{fig:ikuta-model-setup} shows the initial configurations and parameters of a one dimensional hydrodynamic simulation of the flow along the expanding magnetic loop (pseudo 2D MHD model; Section \ref{sec:3-3}) described in \citet{2024ApJ...963...50I} and Ikuta et al. in prep..
The parameter details are described in Section \ref{sec:3-3}.
}

\section{Spatial Resolution of (Zeeman) Doppler Imaging}\label{app:a}

Here we evaluate the spatial resolution of DI and ZDI maps.
The spatial resolution of obtained DI and ZDI maps is given as 
$\delta l = 90 \, {\rm deg} \cdot  \frac{\Delta \lambda }{\lambda} \frac{c}{v\sin i}$ or $360 \, {\rm deg} \cdot \Delta \phi$, where $c$ is the light speed, $v\sin i$ is the stellar projected rotational velocity, and $\Delta \phi$ is the average observing phase coverage difference.
Here, spectral resolution $\frac{\Delta \lambda }{\lambda}$ is $\sim$1/65000, ${v\sin i}$ is 16.4 km s$^{-1}$, and $\Delta \phi$ is 1/16 (see Table \ref{tab:obs}).
Therefore, the spatial resolution can be estimated as $\sim$25.3 or $\sim$22.5 deg.
However, the criterion presented here tends to underestimate the size of the smallest resolved element as it is limited by the resolving power of the spectrograph, while the width of the local line profile should be also the limiting factor for NARVAL.






\bibliography{namekata_EKDra_paper2_ver1}{}
\bibliographystyle{aasjournal}



\end{document}